\documentclass[usenatbib,twocolumn]{aastex63}

\pdfoutput=1
\usepackage[utf8]{inputenc}
\usepackage{multirow}
\usepackage{url}
\usepackage{amsmath}	
\usepackage{amssymb}	
\usepackage{graphicx}
\usepackage{color}
\usepackage{natbib}
\usepackage{hyperref}
\usepackage{enumitem}
\usepackage{ulem}
\setitemize{noitemsep,topsep=0pt,parsep=0pt,partopsep=0pt,leftmargin=*}
\usepackage{times,txfonts}
\usepackage{bm}
\usepackage{mathrsfs}

\usepackage[T1]{fontenc}
\usepackage{ae,aecompl}

\bibpunct{(}{)}{;}{a}{}{,} 

\DeclareUnicodeCharacter{00A0}{ }

\usepackage{listingsutf8}

\newcommand{\unitspace}{\,}

\newcommand{\km}{\ensuremath{\unitspace \mathrm{km}}}

\newcommand{\cm}{\ensuremath{\unitspace \mathrm{cm}}}
\newcommand{\g}{\ensuremath{\unitspace \mathrm{g}}}
\newcommand{\Hz}{\ensuremath{\unitspace \mathrm{Hz}}}

\renewcommand{\sec}{\ensuremath{\unitspace \mathrm{s}}}
\newcommand{\yr}{\ensuremath{\unitspace \mathrm{yr}}}
\newcommand{\Gauss}{\ensuremath{\unitspace \mathrm{G}}}
\newcommand{\Kelvin}{\ensuremath{\unitspace \mathrm{K}}}
\newcommand{\Msun}{\ensuremath{\unitspace \mathrm{M}_{\odot}}}

\def\bfnabla{\nabla}
\def\bfx{{\bf x}}

\def\bfv{{\bf v}}
\def\bfb{{\bf b}}
\def\bfj{{\bf j}}
\def\bff{{\bf f}}
\def\bfk{{\bf k}}
\def\bfB{{\bf B}}
\def\bfft{\tilde{\mathbf{f}}}

\def\bfcalD{\mbox{\boldmath{${\cal D}$}}}

\def\rmD{{\mathrm{D}}}

\def\gsim{\;\rlap{\lower 2.5pt\hbox{$\sim$}}\raise 1.5pt\hbox{$>$}\;}
\def\lsim{\;\rlap{\lower 2.5pt\hbox{$\sim$}}\raise 1.5pt\hbox{$<$}\;}

\usepackage{color}

\makeatletter
\def\fvec#1{\underline{\sbox\tw@{$#1$}\dp\tw@\z@\box\tw@}}
\makeatother

\newcommand{\Ten}[2]{\ensuremath{#1 \times 10^{#2}} }

\shorttitle{Neutron Star Atmosphere--Ocean Dynamics}
\shortauthors{N\"attil\"a et al.}

\begin{document}

\title{Neutron Star Atmosphere--Ocean Dynamics}

\author[0000-0002-3226-4575]{Joonas N\"attil\"a}
\affiliation{Department of Physics, University of Helsinki, Gustaf H\"allstr\"omin katu 2, Helsinki, FI 00014, Finland}
\affiliation{Physics Department and Columbia Astrophysics Laboratory, 
Columbia University, 538 West 120th Street, New York, NY 10027, USA}
\affiliation{Center for Computational Astrophysics, Flatiron Institute, 
162 Fifth Avenue, New York, NY 10010, USA}
\email{email: joonas.nattila@helsinki.fi}

\author[0000-0002-4525-5651]{James Y-K. Cho}
\affiliation{Center for Computational Astrophysics, Flatiron Institute, 
162 Fifth Avenue, New York, NY 10010, USA}
\affiliation{Martin A. Fisher School of Physics, Brandeis University, 
415 South Street, Waltham, MA 02453, USA}

\author[0000-0002-5263-385X]{Jack W. Skinner}
\affiliation{Division of Geological and Planetary Sciences, 
California Institute of Technology, Pasadena, CA 91125, USA}
\affiliation{Martin A. Fisher School of Physics, Brandeis University, 
415 South Street, Waltham, MA 02453, USA}

\author[0000-0002-0491-1210]{Elias R. Most}
\affiliation{TAPIR, MC 350-17, California Institute of Technology, 
Pasadena, CA 91125, USA}
\affiliation{Walter Burke Institute for Theoretical Physics, California Institute of Technology, Pasadena, CA 91125, USA}

\author[0000-0002-7301-3908]{Bart Ripperda}
\affiliation{School of Natural Sciences, Institute for Advanced Study,
Princeton, NJ, 08540, USA}
\affiliation{Center for Computational Astrophysics, Flatiron Institute, 
162 Fifth Avenue, New York, NY 10010, USA}

\begin{abstract}
We analyze the structure and dynamics of the plasma atmospheres and Coulomb-liquid oceans on neutron stars.  
Salient dynamical parameters are identified and their values estimated for the governing set of magnetohydrodynamics equations.  
Neutron star atmospheres and oceans are strongly stratified and, depending on the rotation period, contain a multitude of long-lived vortices (spots) and/or narrow zonal jets (free-shear zones) in the large plasma-beta regime---i.e., $\beta_{\rm p} \gg 1$ (hydrodynamic regime).  
In contrast, when $\beta_{\rm p} \lesssim 1$ (magnetohydrodynamic regime), the flow is dominated by a global lattice of effectively fixed magnetic islands (plasmoids)---without any jets.
Understanding the spatio-temporal variability of dynamic atmospheres and oceans on neutron stars is crucial for interpreting observations of their X-ray emissions.  
\end{abstract}

\keywords{
Neutron stars (1108);
Plasma astrophysics (1261);
Hydrodynamics (1963);
Magnetohydrodynamics (1964);
High energy astrophysics (739);
Astrophysical fluid dynamics (101) 
}

\section{Introduction}

Neutron stars are enveloped by a plasma atmosphere and a Coulomb-liquid 
 ocean \citep[e.g.,][]{haensel2007}.
On the large scales (i.e., $L \gtrsim R/20$, where $L$ is the horizontal length scale and $R$ is the radius of the star), the envelopes are extremely thin, stratified, and inviscid.
Hence, the envelopes are nearly two-dimensional (2D), support (internal and external) gravity waves, and are expected to be turbulent.

Understanding the atmospheric and oceanic flows on neutron stars is 
important for correctly interpreting observations.  
For example, the flows can bias the mass and radius measurements \citep[e.g.,][]{nattila2017}.
Such turbulent envelopes have long been studied in geophysical and planetary fluid dynamics \citep[e.g.,][and references therein]{cho1996,cho1996a, Dritschel2004ANS, scott2010}.  However, studies have yet to be extended to neutron stars. 
Here we begin the investigation of neutron star atmosphere and ocean dynamics, by identifying the salient parameters and estimating their 
values.

Thus far, the focus in neutron star studies has been on understanding 
the thermonuclear (type-I) X-ray bursts  \citep[e.g.,][]{lewin1993, strohmayer2006} and burst oscillations  \citep[e.g.,][]{strohmayer2001, watts2012}.
To this end, theoretical investigations have modeled the burst propagation \citep[e.g.,][]{bildsten1995, cumming2000, spitkovsky2002, cavecchi2013, cavecchi2015, cavecchi2016, cavecchi2019, eiden2020, harpole2021},
convection \citep[e.g.,][]{garcia2018},
linear waves \citep[e.g.,][]{heng2009, klimachkov2016}, 
and oscillation modes \citep[e.g.,][]{heyl2004, lee2004, chambers2020}.
Significantly, all of the phenomena modeled have been treated as a 
perturbation on a uniform and stationary ocean.  
Our aim here is to highlight the inhomogeneous---and especially the 
dynamic---nature of both the atmosphere and ocean on neutron stars.\\

\newpage

\section{Basic Structure}\label{sec:basics}

\begin{table*}[th]
\begin{center}
\begin{tabular}{l l c c c c}
\\
                                      &  & \multicolumn{2}{c}{MSPs}  & \multicolumn{2}{c}{RPs}   \\ 
\hline
Spin period                          & $P$ ($\sec$)                & \multicolumn{2}{c}{ $10^{-3}$}  & \multicolumn{2}{c}{$1$}   \\ 
Magnetic field                       & $B$ ($\Gauss$)              & \multicolumn{2}{c}{ $10^{8}$}  & \multicolumn{2}{c}{$10^{12}$}   \\ 
                                     &                             & Atmos.      & Ocean       & Atmos.          & Ocean \\ 
\hline
Characteristic density& $\rho$ ($\g\cm^{-3}$)       & $1$         & $10^6$      & $1$             & $10^6$ \\
Mean charge number                   & $Z_\mathrm{eff}$            & 1           & 2           & 5               & 26   \\
Scale height                         & $H_p$ ($\cm$)               & $0.8$       & $550$       & $0.1$          & $170$ \\
Thickness                            & $H$ ($\cm$)                 & $3.3$        & $1700$      & $0.4$           & $440$ \\
Kinematic viscosity                  & $\nu$ ($\cm^2\sec^{-1}$)    & $10^1$      & $10^{-1}$   & $10^{-2}$       & $10^{-1}$ \\
Magnetic diffusivity                 & $\eta$ ($\cm^2\sec^{-1}$)   & $10^{4}$    & $10^{3}$    & $10^{14}$       & $10^{3}$ \\
Ion magnetization                    & $\sigma_i$                  & $\lll 1$    & $\lll 1$    & $10^2$          & $10^{-4}$ \\
Electron magnetization               &  $\sigma_e$                 & $10^{-3} $  & $\lll 1$    & $10^5$          & $10^{-1}$ \\
Ion collision time / gyroperiod      & $\tau_i \omega_{ci}$        & $10^{-1}$   & ---         & $10^{1}$        & --- \\
Electron collision time / gyroperiod & $\tau_e \omega_{ce}$        & $10^{0}$    & ---         & $10^{5}$        & --- \\
Gravity wave speed                   & $v_g$ ($\cm\sec^{-1}$)      & \Ten{9}{7}  & \Ten{5}{8}  & \Ten{2}{7}      & \Ten{3}{8} \\
Sound speed                          & $c_s$ ($\cm\sec^{-1}$)      & \Ten{1}{7}  & \Ten{3}{8}  & \Ten{4}{6}      & \Ten{2}{8} \\
Alfv\'en speed                       & $v_A$ ($\cm\sec^{-1}$)      & \Ten{3}{7}  & \Ten{3}{4}  & $\rightarrow c$ & \Ten{3}{8} \\
Selected characteristic speed        & $U$ ($\cm\sec^{-1}$)        & $10^7$      & $10^7$      & $10^7$          & $10^7$ \\
Froude number                        & $F_r$                       & $0.1$       & 0.02        & $0.4$           & $0.04$ \\
Rossby number                        & $R_o$                       & \Ten{7}{-4} & \Ten{7}{-4} & $0.7$           & $0.7$ \\
Burger number                        & $B_u$                       & \Ten{4}{-5} & \Ten{1}{-3} & $3$             & $400$ \\
Brunt--V\"ais\"al\"a frequency       & $N$ ($\sec^{-1}$)           & \Ten{4}{7}  & \Ten{8}{5}  & \Ten{1}{7}      & \Ten{5}{5} \\
Reynolds number                      & $R_e$                       & $10^{12}$   & $10^{14}$   & $10^{15}$       & $10^{14}$ \\
Magnetic Reynolds number             & $R_m$                       & $10^9$      & $10^{10}$   & $0.1$           & $10^{10}$ \\
Plasma-beta                          & $\beta_\mathrm{p}$          & $0.4$       & $400$       & $\lll 1$        & 0.04 \\
Kinematic damping time               & $t_{d,\nu}$ ($\yr$)         & \Ten{5}{4}  & \Ten{5}{5}  & \Ten{5}{6}      & \Ten{5}{5} \\
Magnetic damping time                & $t_{d,\eta}$ ($\yr$)        & $5$         & 50          & $\lll 1$        & 50 \\
Rossby deformation scale    & $L_D$ ($\cm$)               & \Ten{7}{3}  & \Ten{4}{4}  & \Ten{2}{6}      & \Ten{2}{7} \\
Rhines scale                        & $L_\beta$ ($\cm$)           & \Ten{6}{4}  & \Ten{6}{4}  & \Ten{2}{6}      & \Ten{2}{6} \\ [1ex]
\hline\\
\end{tabular}
\caption{Summary of physical quantities and dimensionless numbers in 
the neutron--star atmosphere--ocean layers.
The values are representative; actual parameter values can vary by orders-of-magnitude, depending on the exact details of the layers \citep[see, e.g.,][]{ventura2001} 
}\label{tab:1}
\end{center}
\end{table*}

Neutron stars are typically of a radius and mass,
\begin{equation}
    R\, \approx\, 12\,\km \quad\ \ \mathrm{and} 
    \quad\ \  M\, \approx\, 1.5 \Msun\ ,
\end{equation}
respectively \citep[e.g.,][]{Ozel:2016oaf}.
Given these values, the surface gravity is
\begin{equation}\label{eq:surface_grav}
    g\ =\ \frac{G M}{R^2}\,\left(\frac{1}{\sqrt{1 - r_g/R}}\right)\ \ 
    \approx\ \ \Ten{1.7}{14}\,\cm\,\sec^{-2}\, ,
\end{equation}
where $r_g / R \approx 0.4$, and $r_g \equiv 2G M/c^2$ is the Schwarzschild radius, $G$ is the gravitational constant, and $c$ is the speed of light.
Beyond $R$ and $M$, observed neutron stars can be classified based on their rotation period $P$ and intrinsic (dipole) magnetic field strength $B$ into  
\citep[e.g.,][]{nattila2022c}:
\textit{i}) radio pulsars (RPs), 
\textit{ii}) millisecond pulsars (MSPs), and
\textit{iii}) magnetars.
Here, we focus on the first two classes, as the properties of the magnetar atmosphere and ocean are currently poorly understood.
We take
\begin{align}
      P\, \sim\, 1\,\sec\ \mathrm{,} \qquad \ \ B\, \sim\, 10^{12}\,\Gauss \qquad \ \ {\rm (RPs)} \quad\ \ \ \ \\
      P\,  \sim\, 10^{-3}\,\sec\ \mathrm{,} \quad \  B\, \sim\, 10^8\,\Gauss \qquad\ \ \   {\rm (MSPs)}\ . \quad 
\end{align}

In the upper layers of the atmosphere--ocean layer, the temperature distribution varies from the surface to deeper, nearly isothermal interiors. 
These variations have been thoroughly studied \citep[e.g.,][]{gudmundsson1982, potekhin1997, potekhin2003, potekhin2015}.
For neutron stars without magnetic fields, the temperature distribution is spherically symmetric, $T=T(\rho)$. 
We approximate it by
\begin{align}\label{eq:T}
    T(\rho)\ =\ T_\mathrm{top}\,\left(\frac{ \rho/\rho_{\mathrm{top}}}
    {1 + \rho/\rho_\mathrm{tr}}\right)^{\,\alpha}\, ,
\end{align}
where $T_\mathrm{top}$ is the effective surface temperature of the star at the effective (radiative) surface density $\rho_\mathrm{top}$; 
$\rho_\mathrm{tr}$ is the density at which the electrons become degenerate and radiative (photon) heat conductivity near the surface is replaced by the electron conductivity;
$\alpha$ is a fit parameter. 
For the surface layers composed of iron, we use $\alpha=1/4$ and $\log(\rho_\mathrm{tr}) \approx 5.5 \, T_{\mathrm{top},6}^{1/4}$, where $\rho_\mathrm{tr}$ is expressed in $\g\cm^{-3}$, and $T_{\mathrm{top},6}=T_\mathrm{top}/10^6 \Kelvin$. 
For the layers composed of accreted matter, we use $\alpha=1/3$ and $\log(\rho_\mathrm{tr}) \approx 3.75 \, T_{\mathrm{top},6}^{1/4}$. 
These approximations are in reasonable correspondence with the accurate calculations cited above and have a typical error of $\ \lesssim 20\%\,$ for $T_{\mathrm{top},6} \in [ 0.2, 2.0 ]$.
In the presence of strong magnetic fields, the thermal conductivity in the outer atmosphere-ocean layer becomes anisotropic. 
Then the temperature distribution depends not only on $\rho$ but also on the magnetic field strength and geometry. 
In this way the temperature distribution appears to be nonuniform in a complicated way; 
however, all these effects are not critical for our analysis.

\begin{figure}[t]
\centering
\includegraphics[clip, trim=0.0cm 6.0cm 0.0cm 0.0cm, width=8.5cm]{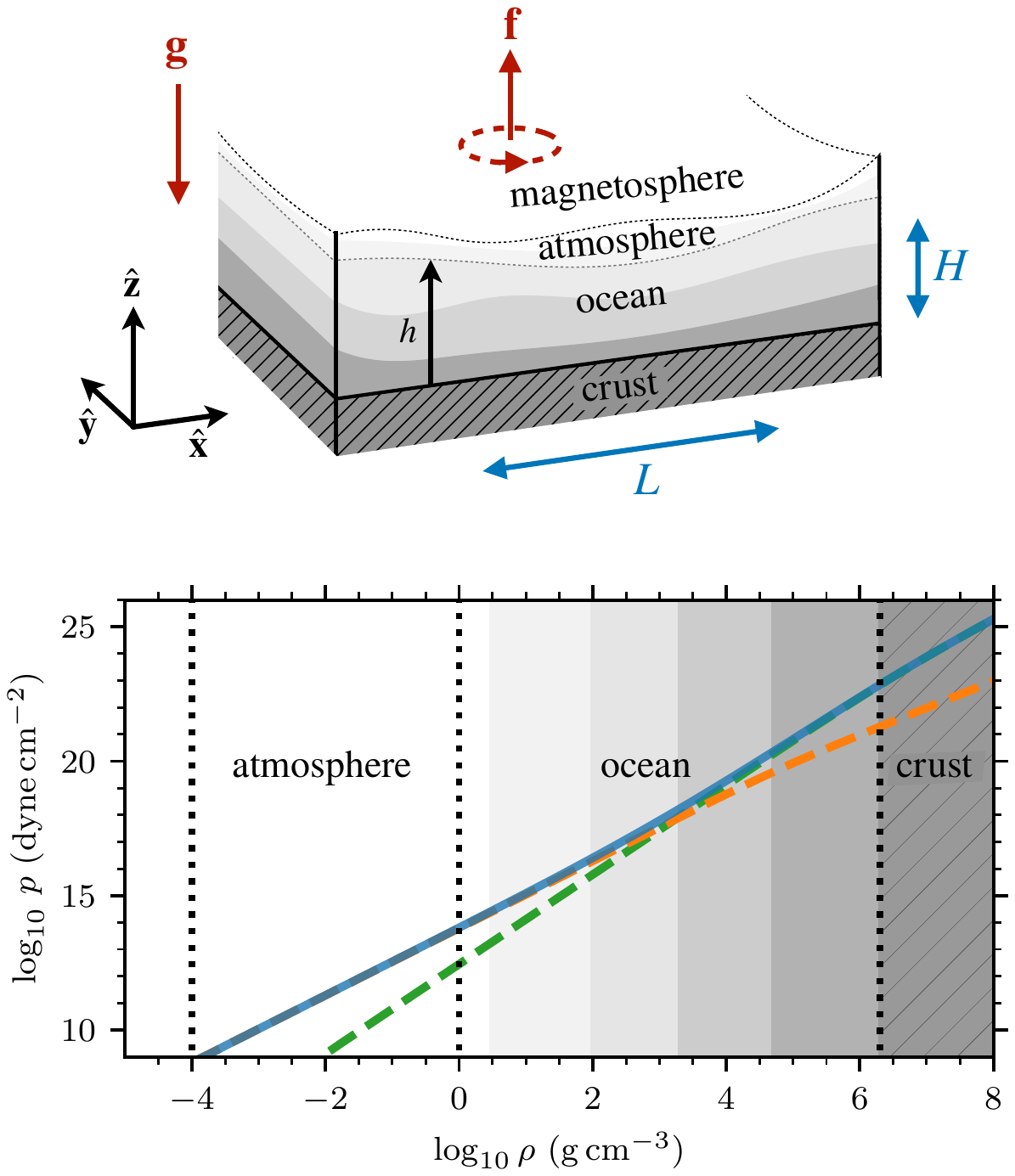}
\caption{\label{fig:layer} Atmosphere--ocean layers.
    Top panel illustrates the geometry of the stratified fluid layers, along  with several important forces and lengths.
    Bottom panel shows the pressure $p$ as a function of density $\rho$ 
    (blue solid curve). 
    The pressure has two main components: an ideal gas law (orange dashed curve) and a degenerate electron gas pressure (green dashed curve).
    Note that the physical height of the atmosphere is much less than that of the ocean (e.g., $\sim$$1\cm$ and $\sim$$10^3\cm$, respectively), as discussed in the text.}
\vspace*{0.5cm}      
\end{figure}

In general, neutron stars have an internal structure consisting of a 
solid (ion) crust and a liquid (neutron) interior \citep{haensel2007}.
Overlying the solid crust are the liquid ocean and the gaseous 
atmosphere \citep[e.g.,][]{gudmundsson1982}.
The latter extends out to the magnetosphere. 
Interestingly, the basic vertical (radial) structure is  
similar to that of the Earth; see Fig.~\ref{fig:layer}.
However, the atmosphere--ocean transition on the Earth is 
sharp, unlike that on the neutron star.
In this paper, the focus is entirely on the atmosphere--ocean regions of RPs and MSPs.
Table~\ref{tab:1} summarizes physical quantities and dimensionless numbers for the regions.

\subsection{Atmosphere}

The equation of state for the plasma atmosphere is aptly described by 
the ideal gas law: 
$p = n k_\mathrm{B} T = \rho \mathcal{R} T$, where 
$p$ is the pressure,
$n$ is the number density of ions and electrons,
$k_\mathrm{B}$ is the Boltzmann constant, 
$T$ is the temperature, 
$\rho$ is the density, and 
$\mathcal{R}$ is the specific gas constant (Fig.~\ref{fig:layer}).
Note, the atmosphere is generally baroclinic.
The exact definition of the top and bottom boundaries for the atmosphere are somewhat arbitrary.  
However, it is natural to define the atmosphere so that it encapsulates the star's ``photosphere'' (where the optical depth of an X-ray is approximately unity) at 
$\rho \sim 10^{-2}\,\g\cm^{-3}$ \citep[e.g.,][]{potekhin1997}.
We further take the ``top of the atmosphere'' to be at $\rho = 10^{-4}\,\g\cm^{-3} \equiv \rho_\mathrm{top}$\,, since the plasma becomes strongly magnetized at
$\rho \lesssim 10^{-4}\,\g\cm^{-3}$.
As for the bottom, the gaseous atmosphere transitions to the ocean where the equation of state starts to be modified by Coulomb forces, at $\rho \sim 1\,\g\cm^{-3} \equiv \rho_\mathrm{tr}$.

The transport properties of the atmosphere are mainly mediated by Coulomb collisions.
The collision times (needed for viscosity, for example) at $\rho = 1\,\g\cm^{-3}$ can be estimated following  \citet{spitzer1962}: 
for a plasma with an effective charge number $Z \approx 5$, mass number $A = 56$, and $T = \Ten{5}{6}\,\Kelvin$ \citep[e.g.,][]{potekhin1997}, the collision times for the electron and the ion ($\tau_e$ and $\tau_i$, respectively) are approximately equal---i.e., $\tau_e \sim \Ten{3}{-15} \sec$ and $\tau_i \sim \Ten{5}{-15}\sec$.
However, strong magnetic fields characteristic of RPs can ``magnetize'' the atmospheric electrons and ions so that $\tau_e\,\omega_{ce} \gg 1$ and $\tau_i\,\omega_{ci} \gg 1$, where $\omega_{ce}$ is the electron gyrofrequency and $\omega_{ci}$ is the ion gyrofrequency; 
in this case, the isotropic viscosity and resistivity estimates are not valid \citep[][]{braginskii1965}.
At stronger magnetic fields, the transport properties of electrons in the atmosphere and ocean can also be affected by Landau quantization of electron motion. 
However, these effects are unimportant for our study.

For weaker magnetic fields, which are characteristic of MSPs, the ions remain un-magnetized throughout the atmosphere while the electrons become magnetized in the upper region of the atmosphere ($\rho \ll 1\, \g \cm^{-3}$).
The ion collisions dissipate momentum with a dynamic viscosity of $\mu \approx 0.96\,\rho v_{\mathrm{th}}^2\tau_i \sim 10^{-2}\, \g\cm^{-1}\sec^{-1}$, where $v_{\mathrm{th}} \equiv (k_\mathrm{B} T / m_i)^{1/2}$ is the thermal speed of the ion.
Hence, the kinematic viscosity, $\nu\ \equiv\ \mu / \rho$, is $\,\sim$$1\,\cm^2 \sec^{-1}$ for both RPs and MSPs. 
Similarly, the electron collisions dissipate the electric current with a perpendicular Spitzer conductivity, $\sigma_\perp \equiv n e^2/m_e \omega_{ce}^2 \tau_e$, of $\sim$$10^{15} \sec^{-1}$ for $B = 10^{8}\Gauss$ (MSPs) and $\sim$$10^6\sec^{-1}$ for $B = 10^{12}\Gauss$ (RPs).
Then, the corresponding magnetic diffusivity, $\eta \equiv c^2/4\pi \sigma_\perp$, is $\sim$$10^{4} \cm^2 \sec^{-1}$ for MSPs and $\sim$$10^{14} \cm^2 \sec^{-1}$ for RPs.

\subsection{Ocean}

The equation of state for the ocean is described by the equation for a degenerate electron gas, $p \propto \rho^\gamma$\,; 
here $\gamma \in [4/3,\,5/3]$, depending on the region of the ocean.
The equation of state for the upper region of the ocean ($\rho \lesssim 10^6\,\g\cm^{-3}$) is described by the equation for the non-relativistic, degenerate electron gas, $\gamma = 5/3$.
Below this region, the electrons are relativistic. 
Hence, the equation of state for the lower region is described by that 
for the relativistic, degenerate electron gas, $\gamma = 4/3$. 
Degeneracy pressures in both regions of the ocean do not have a strong temperature dependence.  
Therefore, unlike the atmosphere, the ocean is wholly barotropic; 
more specifically, it is polytropic.

The bottom of the ocean is defined to be located where the 
Coulomb liquid freezes into a crystal lattice.
The latter can be adequately modeled as a plasma with ions of one type \citep[see, e.g.,][]{chamel2008}.
For simplicity, we assume here a canonical composition of fully-ionized Fe ($Z = 26$ and $A = 56$).
The liquid-to-solid transition is characterized by the ratio of the electrostatic interaction energy to the thermal energy.
This is quantified by the Coulomb-coupling parameter \citep[e.g.,][]{chamel2008},
\begin{equation}
    \Gamma_\mathrm{C}\ \equiv\ \frac{Z^2 e^2}{a_i k_B T}\ ,
\end{equation}
where $e$ is the electron charge and $a_i = (4\pi n_i/3)^{-1/3}$ is the ion sphere radius with $n_i$ the number density of ions.  
The melting of a homogeneous Coulomb lattice occurs at $\Gamma_{\mathrm{C}} \approx 175$ \citep{potekhin2000}.
This corresponds to $\,\rho_\mathrm{bot} \approx \Ten{2}{6}\,\g\cm^{-3}$, given the $T(\rho)$ of Eq.~\eqref{eq:T}. 
The exact value of $\Gamma_{\mathrm{C}}$ and profile of $T$---hence $\,\rho_\mathrm{bot}$---can differ depending on the plasma composition, 
lattice impurities, external heating processes, and magnetic 
fields.
However, this uncertainty does not qualitatively change the basic 
picture presented here. 

The transport properties of a Coulomb liquid are mainly mediated by 
electrons \citep{chamel2008}.
Therefore, the momentum dissipation in the ocean is expected to be 
dominated by the electron shear viscosity, 
$\mu\approx \Ten{6}{4}\,\g\cm^{-1} \sec^{-1}$ \citep[][calculated 
for $T = 10^7 \Kelvin$]{chugunov2005}. 
We ignore the bulk viscosity since the flow discussed here is nearly incompressible (i.e., characteristic flow speed is small compared to 
the sound speed).
This gives $\nu \sim 0.1\, \cm^{2}\sec^{-1}$.
The electrical conductivity in the ocean is strongly anisotropic with 
respect to the direction of the local magnetic field: 
the longitudinal conductivity~$\sigma_\parallel$ is $\sim$$\Ten{3}{20}\,\sec^{-1}$, and the transverse conductivity~$\sigma_\perp$ is $\sim$$\Ten{6}{16}\,\sec^{-1}$ (\citealt{potekhin1999, potekhin2003}; calculated for $T = 10^7\,\Kelvin$ and $B = 10^{12}\,\Gauss$).
The magnetic diffusivity (using the smaller conductivity), 
$\eta \equiv c^2/4\pi \sigma_\perp$, is $\sim$$10^3\,\cm^{2}\sec^{-1}$.
This is a small value for the dynamics on the large scale (see Section~\ref{sec:dynamics}); hence, transport properties do not have a significant, direct influence on the dynamics at that scale.

\subsection{Thinness of the Upper Layers}

The atmosphere-ocean layer forms a thin fluid shell on top of the solid crust.
The layer's height can be quantified by the pressure scale height,
$H_p = H_p\,(\rho, T)$.
Under the hydrostatic balance condition, which is appropriate for 
the large scales, it is simply
\begin{equation}\label{eq:h_p_atmos}
    H_p\ =\ \frac{p}{g\rho} \, 
\end{equation}
for the atmosphere, if $T \ne T(\rho)$.
This gives $H_p \approx 0.1 \cm$ for RPs and $\approx\! 0.8\cm$ for MSPs at $\rho = 1 \g \cm^{-3}$.
On the other hand, the ocean scale-height is given by
\begin{equation}\label{eq:h_p}
    H_p\ \approx\ \left[ e^{(\gamma - 1)/\gamma} - 1 \right]\,
    \frac{\gamma}{\gamma - 1}\ \frac{p}{g \rho}\, .
\end{equation}
Using $\gamma = 5/3$, this gives $H_p \approx 170\cm$ for RPs 
and $\approx\! 550\cm$ for MSPs at $\rho = 10^6\,\g\cm^{-3}$.
For both the atmosphere and ocean, the physical thickness generally 
spans a large number of scale heights---e.g., the number is 
$\,\gtrsim\! 20$ for the Earth's atmosphere and ocean.  
For neutron stars, the thickness of the atmosphere (corresponding to $\rho_\mathrm{top} \le \rho \le \rho_\mathrm{tr}$) is $\sim$$4\, H_p$---i.e., $\sim$$0.4\cm$ for RPs and $\sim$$3\cm$ for MSPs.
The thickness of the ocean (corresponding to $\rho_\mathrm{tr} \le \rho \le \rho_\mathrm{bot}$) is $\sim$$3\,H_p$---i.e., $\sim$$440\cm$ for RPs and $\sim$$1700\cm$ for MSPs.
Therefore, the bulk of the combined atmosphere--ocean thickness ($\sim$$10\, H_p$) is taken up by the ocean.
In the subsequent analysis, we take the characteristic thickness of the modeled layer $H$ $(\sim\! H_p)$ to be $500\cm$ for both RPs and MSPs.

\newpage
\section{Dynamics Analysis}\label{sec:dynamics}

\subsection{Governing Equations}
The large-scale dynamics of the electrically neutral (un-ionized) atmosphere and ocean are governed by the primitive equations 
\citep[e.g.,][]{pedlosky1979}, 
which can be thought of as describing a system of stacked thin-layers 
of fluid \citep[e.g.,][]{ripa1993}.
Each layer, in turn, can be described by the shallow-water equations 
(SWE) for homogeneous fluid 
\citep[e.g.,][]{pedlosky1979,gill1982}---provided that the vertical 
to horizontal aspect ratio of the flow structures in the layer is 
small.%
\footnote{
The condition on the aspect ratio is akin to the condition of hydrostatic balance, which strongly decouples the vertical dynamics from the lateral dynamics; and, the homogeneous  
condition refers to the lack of lateral density variation, e.g.,  
which fully takes the thermodynamics out of the system.
}
Then, the starting point of our analysis is the ``magnetized'' form 
of the SWE \citep{gilman2000}, the magnetohydrodynamic shallow-water 
equations (MHDSWE):
\setlength{\jot}{8pt}
\begin{align}
    \frac{\rmD \bfv}{\rmD t}\ & =\ -\frac{1}{F_{\! r}^{\,2}}\bfnabla 
    h - \frac{1}{R_o}\bfk \times \bfv  + 
    \frac{1}{\beta_{\! p}^{\ 2}}\bfb
    \cdot \bfnabla\,\bfb  + \frac{1}{R_e}\bfcalD_\bfv \quad  \label{eq:mhdswe_momentum}\\
    \frac{\rmD h}{\rmD t}\ & =\ -h\,\bfnabla \cdot \bfv  \label{eq:mhdswe_height}\\
    \frac{\rmD \bfb}{\rmD t\,}\, & =\ \bfb \cdot \bfnabla\,\bfv + 
    \frac{1}{R_m}\bfcalD_\bfb\, ,  \label{eq:mhdswe_induction}
\end{align}
where
$\rmD/\rmD t\ \equiv\ \partial/\partial t + \bfv \cdot \bfnabla$ is the 
material derivative,
$\bfv = \bfv(\bfx, t) \equiv (v_x, v_y)$ is the velocity,
$\bfb = \bfb(\bfx, t) \equiv (b_x, b_y)$ is the magnetic field,
$h = h(\bfx, t)$ is the free-surface height, and
$\bfk$ is a unit vector normal to $\bfx$, where $\bfx, \bfnabla \in \mathbb{R}^2$; hence, $\bfv \ne \bfv(z)$ and $\bfb \ne \bfb(z)$, where 
$z$ is the  coordinate along $\bfk$.  
The terms containing $\bfcalD_\bfv$ and $\bfcalD_\bfb$ represent viscous and resistive dissipations, respectively; for example, $\bfcalD_\bfv \rightarrow \bfnabla^2\,\bfv$ and $\bfcalD_\bfb \rightarrow \bfnabla^2\,\bfb$, when $\bfnabla\cdot\bfv \rightarrow 0$.
Eqs.~(\ref{eq:mhdswe_momentum})--(\ref{eq:mhdswe_induction}) are dimensionless.
The dimensionless parameters therein,
\{$F_r$, $R_o$, $\beta_p$, $R_e$, $R_m$\}, follow from the below scaling (see also Appendix~\ref{app:mhdswe}):
\begin{align}
    \bfnabla & \rightarrow \bfnabla\,/\,L\, , \quad 
    t \rightarrow  (L/U)\,t\, , \nonumber \\
    \bfv & \rightarrow U\,\bfv\, , \quad\ \, h \rightarrow H\,h\, , 
    \qquad\ \bfb \rightarrow \bfb\,/\!\sqrt{4\pi \rho}\ ,
\end{align}   
where $U$ and $\rho$  are the characteristic speed (of the structure of size $L$) and density, respectively. 

\subsection{Forcing Mechanisms}\label{sect:forcing}

Atmospheric and oceanic flows on neutron stars can be readily forced by one or more of the following mechanisms or perturbations: 
\begin{itemize}[topsep=0pt,itemsep=0pt,leftmargin=2\parindent]
    \item[{\it i})] shear and dilatation of the crustal layers, due to star quakes \citep{anderson1975}
    
    \item[{\it ii})] local heat flux from electron--positron pair 
    cascades in the magnetosphere \citep{sturrock1971}
    
    \item[{\it iii})] gravity waves generated by $\bfb$-field line 
    perturbations \citep{blaes1989}
    
    \item[{\it iv})] oblateness in fast rotators, which present ``stellar-scale topography'' that induce poloidal flows as well as excite gravity waves---e.g., when jets flow over 
     it \citep{algendy2014}
    
    \item[{\it v})] ``dayside--nightside'' thermal as well as 
    gravitational forcing in close-in binary systems and during 
    actual mergers \citep{sullivan2022}
    
    \item[{\it vi})] thermonuclear burning of the ocean material and moving burning front  \citep{spitkovsky2002}
    
    \item[{\it vii})] ``primordial'' perturbation from the protoneutron star collapse \citep{burrows1986} 

    \item[{\it viii})] boundary/spreading layer in low-mass X-ray binary systems with an accretion 
    disk \citep{inogamov1999,inogamov2010, abolmasov2020}
    
    \item[{\it ix})] oceanic flow induced by atmospheric winds, and 
    their back-reaction \citep[e.g.,][]{gill1982}
    
    \item[{\it x})] magnetically channeled matter into the star's poles from a truncated disk \citep[e.g.,][]{abolmasov2022}.
\end{itemize}
The above list is inexhaustive. 
When $\beta_p$ is finite, the $\bfb\cdot\bfnabla\,\bfb$ 
term in Eq.~(\ref{eq:mhdswe_momentum}) presents a source of 
momentum---even in the absence of motion initially.
Once driven into motion, the layers typically evolve through one or 
more quasi-equilibrium flow states, each of which can be roughly 
understood as balances between various terms in 
Eqs.~(\ref{eq:mhdswe_momentum})--(\ref{eq:mhdswe_induction}).

\subsection{Dynamical Parameters}

In this section, we discuss parameters important for the characteristic flow speed, stratification, rotation, and magnetic field strength. 
The parameters provide useful information even without explicitly 
solving the equations.
The MHDSWE are divergent in the lateral direction---i.e., 
$\bfnabla \cdot \bfv \ne 0$; hence, (external) gravity waves are admitted. 
These are edge waves that propagate with speed,
\begin{equation}
    v_g\ \equiv\ \sqrt{g H}\ \ \approx\ \ \Ten{3}{8}\,\cm\sec^{-1}\, .
\end{equation}
However, the MHDSWE are derived assuming non-divergence in three dimensions.
Formally, this means $c_s \rightarrow \infty$, where $c_s$ is the sound speed; 
hence, acoustic waves are filtered out and thermodynamics is 
decoupled from the system.   
If the preceding assumption is not made, $c_s$ (always $\ge v_g$) presents a practical upper limit for the 
flow speed. 
In the adiabatic case,
\begin{equation}
    c_s\ =\ \sqrt{\frac{\partial P}{\partial \rho} }\ =\ 
    \sqrt{\gamma\,\frac{P}{\rho} }\ \
    \approx\ \ \Ten{2}{8}\,\cm\sec^{-1}\, .
\end{equation}
As for the value of the characteristic flow speed $U$, noting that typically $U \ll c_s$ \citep[e.g.,][]{cho2008,cho2019}%
\footnote{
For the Sun and all the other objects that have been spatially resolved (e.g., Solar System planets and Titan), $U \ll v_g$---except for the extremely cold ($T\!\sim\! 50$\,K) gas giant planets, for 
which $U \lesssim c_s/2$; in general, $U/c_s$ decreases as the $T$ increases.
}%
, we choose $U = 10^{7} \cm\sec^{-1}$.

By construction, the above choice of $U$ sets several crucial 
non-dimensional numbers.  For example, the Froude number,
\begin{equation}
    F_r\ \equiv\ \frac{U}{\sqrt{g H}}\, ,
\end{equation}
is the analog of the Mach number for systems in which $c_s \rightarrow \infty$; 
here $F_r \sim 10^{-2}$.
Since $F_r < 1$, the layer described by the MHDSWE remains subsonic. 
Flow structures in a $F_r \nll 1$ system are less coherent (organized) 
in their appearance in the physical space, compared to those in a 
$F_r \ll 1$ system.

The importance of the neutron star's rotation is given by the Rossby 
number,
\begin{equation}
    R_o\ \equiv\ \frac{U}{f L}\, ,
\end{equation}
which measures the relative strengths of the vorticity of the flow 
structure (of size $L$) to the stellar spin rate $\Omega \equiv 2\pi/P$ (the stellar vorticity) via the Coriolis parameter, $f \equiv 2 \Omega \sin\theta$.
For RPs, $R_o \sim 1$ for $L \sim R$.
In contrast, $R_o \sim 10^{-3}$ for MSPs, indicating that structures 
of $L \sim R$ are rotationally dominated.%
\footnote{%
One implication of this is that, for RPs, reduced systems based on small $R_o$, such as the quasi-geostrophic equations \citep[][]{pedlosky1979}, cannot be used.  
For MSPs, the cruciality of retaining the gravity waves in the system dictates whether the quasi-geostrophic equations remain valid, as the waves are filtered out from the equations.
}

The distance a gravity wave effectively traverses in one stellar 
rotation is characterized by the (external) Rossby deformation scale,
\begin{equation}\label{eq:deform}
    L_D\ \equiv\ \frac{\sqrt{gH}}{f}\, .
\end{equation}
In the context of shallow-water models, $L_D$ plays the role of an interaction length between flow structures.
Note that $R_o$ and $L_D$---as well as other quantities below---are 
dependent on the latitude $\theta$, through~$f$.   
From hereon, we take the mid-latitude value of $f$ as its characteristic value, $f_0 = \Omega\sqrt{2}$. 
Then, for MSPs, $L_D \approx \Ten{2}{4} \cm$, which is $\sim\! R/100$; 
and, for RPs, $L_D \approx \Ten{2}{7} \cm$, which is $\sim\! 10\,R$. 
For the MSPs, the large scales then naturally divide into two sub-scales: 
a {\it stellar} scale ($L \gtrsim R$) and a {\it synoptic} scale 
($R \gtrsim L \gtrsim L_D$).  
For the RPs, given their size, $L_D$ (hence gravity waves) plays a relatively minor role in the large-scale dynamics---analogous to the filtering of acoustic waves discussed above.  

With $R_o$ and $F_r$ identified, another important non-dimensional parameter that naturally emerges is the Burger number,
\begin{equation}
    B_u\ \equiv\ \left(\frac{R_o}{F_r}\right)^2\ =\ 
    \left( \frac{L_D}{L} \right)^2\, .
\end{equation}
It is a measure of the interaction length in terms of the characteristic length scale.
For MSPs, $B_u$ tends to be small, $\sim$$10^{-3}$, throughout the surface layers. 
In contrast, for RPs, $B_u >1$, as it ranges from $\sim$$3$ in the atmosphere to $\sim 400$ in the ocean.
As indicated, $B_u$ is also a measure of $L_D$ in terms of~$L$.  
Under full stratification (i.e., baroclinic, requiring multiple layers), 
\begin{equation}\label{eq:Bu}
B_u\ =\ \left(\frac{NH}{f\,L}\right)^2\, , 
\end{equation}
where $N$ is the local Brunt--V\"ais\"al\"a 
frequency%
\footnote{%
Under general stratification $z = z(\bfx,t)$; 
also, from Eq.~\eqref{eq:Bu}, we obtain $L_D = N H/f < (gH)^{1/2} / f$,
sometimes referred to as the {\it internal} $L_D$ to distinguish
it from the external $L_D$.%
}%
\,:
\begin{equation}
    N\ \equiv\ 
    \sqrt{\frac{g^*}{\Xi}\,\frac{\partial \Xi}{\partial z}}\ ,
\end{equation}
where $\Xi = \Xi\,(\bfx,z,t)$ is the potential density.
Under the hydrostatic balance condition, $\Xi \approx \Theta(z)$ with $g^* = g$ for the baroclinic (atmosphere) region, and $\Xi \approx -\rho(z)$ with $g^* = \gamma\,(\gamma - 1)\,g$ for the 
barotropic (ocean) region.
Here $\Theta \equiv T(p_R/p)^\kappa$ is the potential temperature, where $p_R$ is a constant reference $p$-value, $\kappa \equiv {\cal R}/\mathfrak{c}_p = \gamma - 1$, and $\mathfrak{c}_p$ is the specific 
heat at constant $p$. 
The lateral variation $\Theta$ is assumed to be small compared to the vertical variation in $\Theta$ and $\rho$.
We also note here that $\Theta$ is related to the specific entropy $s$ by ${\rm d}s = c_p\,{\rm d}(\ln\Theta)$; 
hence, $N$ is a measure of the vertical entropy gradient under general baroclinic, thermodynamically reversible conditions---including for the ocean, where $\alpha = \kappa$.
In general, $N \sim 10^6$ to $10^7\, \Hz$ at the atmosphere--ocean layers of neutron stars.
This gives a Prandtl ratio, 
\begin{equation}
    P_r\ \equiv\ \frac{N^2}{\Omega^2}\, ,
\end{equation} 
of $\gtrsim 10^{10}$ for the RPs and $\gtrsim 10^{4}$ for the MSPs.
Its large value signifies that atmospheres and oceans on neutron stars are, in general, extremely stratified (cf., $P_r \sim 10^4$ for the Earth's troposphere).
Given the discussion thus far, two natural timescales can be identified in the system: a short one ($2\pi/N$) and a long one ($2\pi/\Omega$).

The relative strength of the nonlinear advection term to the viscous 
term in Eq.~(\ref{eq:mhdswe_momentum}) is quantified by the Reynolds 
number,
\begin{equation}
    R_e\ \equiv\ \frac{U L}{\nu}\, ,
\end{equation}
where $\nu \sim 0.1$-$1 \cm^2\sec^{-1}$ is the kinematic viscosity. 
For the scales considered here, $R_e$ is exceedingly large.
For example, $R_e \sim 10^{11}$ with $L \sim R$ for the atmospheres of 
both RPs and MSPs; it is larger still for the oceans.
Therefore, viscosity can be ignored.
Correspondingly, the large $R_e$ value gives a very long turbulent damping time:
\begin{equation}
    t_{\mathrm{d},\nu}\ \equiv\ \frac{R^2}{\nu}\ \ 
    \sim\ \  2\times 10^{12} \sec\, ,
\end{equation}
indicating that---once perturbed---the fluid on the large scale is 
effectively always in motion, in the absence of cooling.

Similarly, the relative strength of the resistive term to the nonlinear 
advection term in Eq.~(\ref{eq:mhdswe_induction}) is quantified by the 
magnetic Reynolds number,
\begin{equation}
    R_m\ \equiv\ \frac{U L}{\eta}\, ,
\end{equation}
where $\eta$ is the magnetic diffusivity.
With $L \sim R$, this number is large for the ocean of RPs and MSPs: 
$R_m \gtrsim 10^{10}$ for $\eta \sim 10^{3} \cm^2\sec^{-1}$.
Hence, resistivity can also be ignored for the large-scale ocean flows.
Accordingly, the turbulent resistive time is large:
\begin{equation}
    t_{\mathrm{d},\eta}\ \equiv\ \frac{R^2}{\eta}\ \ 
    \sim\ \ 2\times10^8 \sec\, ,
\end{equation}
indicating that---once perturbed---the $\bfb$-field of the oceanic fluid is also effectively always perturbed.

The weakly-magnetized atmospheric layers of MSPs have a similarly large $R_m \sim 10^{12}$ for $\eta \sim 10^{4} \cm^2\sec^{-1}$.
In contrast, the strongly magnetized atmospheric layers of RPs have a small $R_m \sim 0.1$ (for $\eta \sim 10^{14} \cm^2\sec^{-1}$) and a short turbulent resistive time, due to the large resistivity of the plasma.
This presents a resistive over-layer for the ocean of RPs, providing large magnetic damping for flow structures (e.g., vertically propagating magneto-gravity waves).
The magnetic Prandtl number, 
\begin{equation}
    P_m\ \equiv\ \frac{R_m}{R_e}\, ,
\end{equation} 
is a measure of the relative strength of viscosity to resistivity. 
It varies from $\sim$$10^{-16}$ to $\sim$$10^{-4}$ for the atmosphere 
and ocean, respectively, on RPs; 
it is $\,\sim$$10^{-4}$, for both the atmosphere and ocean on MSPs.

As already discussed, the strength of coupling between the $\bfb$-field and the flow is characterized by the plasma-beta parameter,
\begin{equation}\label{eq:plasma_beta}
    \beta_p\ \equiv\ \frac{U}{v_{\mathrm{A}}}\, ,
\end{equation}
where $U$ is used instead of the usual pressure-related speed; here $v_\mathrm{A}$ is the Alfv\'en speed,
\begin{equation}
v_\mathrm{A}\ \equiv\ \frac{b}{\sqrt{4\pi \rho}}\, ,
\end{equation}
which characterizes the propagation speed of magnetic perturbations on $\bfb$-field lines, and $b$ is the characteristic strength of 
$\bfb$.%
\footnote{%
The fastest wave permitted in the MHDSWE is the magneto-gravity wave that propagates with a velocity of $v_{\mathrm{mg}} = (v_g^2 + v_\mathrm{A}^2)^{1/2}$.
However, $v_\mathrm{mg}$ cannot exceed $c_\mathrm{s}$, due to the shallow-water assumption.  
This sets the minimum plasma-beta, $\beta_{p}^\mathrm{\, min} = U/c_s$.
For our assumption of $U \sim 0.1\, c_s$, this would lead to $\beta_{p}^\mathrm{\, min} \sim 0.1$.%
}
As defined, $\beta_p$ is the magnetized layer's analog of $F_r$.
We emphasize that $\beta_p$, as defined in Eq.~(\ref{eq:plasma_beta}), 
is generally much smaller than the traditionally defined $\beta_p$.

Magnetic fields, both external and internal to the fluid, permeate the atmosphere and ocean layers. 
The ``external'' stellar-field $\bfB$, presumed to be generated and maintained below the crust, threads the layers at various angles from above and below at different locations.
For example, if $\bfB$ is associated with a simple dipole which is oriented close to the rotation axis, the $\bfB$-lines within the layers are predominantly vertical near the poles and horizontal near the equator.  
Additionally, the layers also possess an ``internal'' magnetic field $\bfb$, that is fully encapsulated within the layer \citep[e.g.,][]{gilman2000,cho2008,dritschel2023}.
At present, $b$ is not known. 
 
We estimate $v_\mathrm{A}$ using $B$.
For $\beta_p \gtrsim 1$, we can estimate $v_\mathrm{A}$ using $B$ because $b$ is at most $B$.
As for when $\beta_p \ll 1$, we have that $b$ is at least $B$ or larger.
However, quasi-equipartition of the kinetic and magnetic energies has been observed in some past MHD studies 
\citep[e.g.,][and references therein]{kraichnan1965}, suggesting a relaxation of the flow into $\beta_p \sim 1$.
In general, $\beta_p \lesssim 10$ for neutron star atmospheres.%
\footnote{%
In reality, in the upper atmospheric region of the RPs (only), $v_\mathrm{A} \rightarrow c$.
Then, a more relevant characterization of the magnetic field strength is the magnetization parameter $\sigma$: 
the atmospheres of RPs are magnetically dominated with ion and electron magnetizations of 
  $\sigma_i \equiv B^2/8\pi n_i A m_p c^2\,\sim\, 10^{2}$ 
  and  $\sigma_e \equiv B^2/8\pi n_e m_e c^2\,\sim\, 10^{5}$, 
  respectively.%
}
In contrast,, deep in the ocean (taking $\rho = 10^6\g\cm^{-3}$) $v_\mathrm{A} \sim \Ten{3}{8} \cm\sec^{-1}$ for RPs and $v_A \sim \Ten{3}{4} \cm\sec^{-1}$ for MSPs.
For the full range of neutron stars considered here, this leads to 
a large range of $\beta_p$ for the oceans, from $\sim$$0.04$ to 
$\sim$$400$.
Therefore, when $B \gtrsim b$, the hydrodynamic (HD) and MHD regimes of the oceanic flow (i.e., $\beta_p \gg 1$ and $\beta_p \lesssim 1$, 
respectively) are delineated by the critical $B$ value, 
$B_\mathrm{crit}\sim 10^{10} \Gauss$.

\newpage
\subsection{Flows in Hydrodynamic Regime}

In this section, we discuss several physical processes important for neutron star atmosphere--ocean dynamics when 
$\beta_p, R_e, R_m \rightarrow \infty$. 
In this regime, the flow is HD and turbulent.
Here, if the flow is quasi-2D (i.e., $\bfnabla\cdot\bfv \ll 1$),
random stirring at a small scale leads to a dual cascade---predominant inverse-cascade of energy to large scales and forward-cascade of enstrophy to small scales \citep{kraichnan1967, leith1968, batchelor1969}.
In contrast, in a shallow-water layer (for which $\bfnabla\cdot\bfv \nll 1$, in general), energy can forward-cascade in addition to inverse cascade; 
hence, when perturbed at a large scale, energy is also effectively transferred to small scales.
As a result, motions on a broad range of scales are produced
if the layer is perturbed---either continuously or initially, given 
the large damping time---at any scale away from the dissipation scale $\ell_\nu$ ($\ll L, L_D$). 
The above cascades are a direct consequence of the material conservation,
\begin{equation}
    \frac{{\rm D}\ }{{\rm D} t}\,\Big(\,\mathscr{F}[q]\,\Big)\ 
    =\ 0\, ,
\end{equation}
when $\beta_p, R_e, R_m \rightarrow \infty$.
Here $\mathscr{F}$ is a functional and is any function of $q$, 
including the potential enstrophy $\frac{1}{2} q^2$ and potential 
palinstrophy $\frac{1}{2}q^4$.

During the turbulent evolution, vortices continuously undergo mergers, filamentations, and splitting (see, e.g., \citealt{dritschel1988, polvani1989} and Appendix~\ref{app:vortex}).
In general, the main vortices are elliptically shaped, due to the strain caused by the background flow.  
In the latter, there are two hyperbolic critical points, close to the the major axes of each vortex.
In addition, if the boundary of the vortex is sharp (i.e., there is a vorticity jump across the boundary line), Kelvin (edge) waves are 
supported along the boundary.
Then, as the vortex rotates, the wave encounters a critical point,
which locally distort the boundary into narrow filaments.
When the filament is thin enough, dissipation disconnects it or 
reconnects it with nearby vortices or filaments; the filament can also roll up into additional small-scale vortices.

In quasi-2D turbulence, the preceding filamentation process is traditionally associated with the forward cascade.
The inverse cascade, on the other hand, is traditionally associated 
with vortex mergers (e.g., \citealt{mcwilliams1984}):
two similarly-sized vortices undergoing a merger, for example, generally leads to a shift of kinetic energy to lower wave numbers (larger scales).
However, conservation of potential enstrophy implies that mergers
must also be accompanied by filamentary vorticity and other 
small-scale vorticity debris, which contributes to the forward 
cascade \citep[see, e.g.,][]{reinaud2018}.
For the merger process, it is worth noting that the vortices have a natural shielding length, characterized by $L_D$.
This arises from the following relationship, which is derived from the SWE under the $\bfnabla \cdot \bfv \ll 1$ and $R_o \ll 1$ condition:
\begin{equation}
    \left[\,\nabla^2 + \frac{1}{L_D^2}\,\right]\,\psi\ =\ \ \zeta\, ,
\end{equation}
where $\psi = \psi(\bfx,t)$ is the velocity streamfunction; $\psi$ is defined by $\bfv  = \nabla^\perp\,\psi$, where $\nabla^\perp \equiv (-\partial/\partial y,\,\partial/\partial x)$.
This gives a Green's function kernel ${\cal G}\,(\bfx^\prime;\bfx)$ for the inversion, $\zeta(\bfx) = 
\int\,{\cal G}(\bfx^\prime;\bfx)\,\psi(\bfx^\prime)\ {\rm d}\bfx^\prime$, 
which asymptotically decays in space as $1/L_D$ \citep[e.g.,][]{dritschel1993b}, analogous to the Debye length for plasmas. 
Note that the pure 2D case, in which $\bfnabla\cdot\bfv = 0$, is 
recovered when $L_D \rightarrow \infty$.

The inverse cascade is anisotropically impeded by Rossby wave generation at large scales where the nonlinear advection term can balance with the Coriolis term.
Known as the ``$\beta$-effect'', the balance occurs roughly at the Rhines scale \citep{rhines1975}, 
\begin{equation}
    L_\beta\ \equiv\ \sqrt{\frac{2U}{\beta} }\ \ \sim\ \
    \sqrt{\frac{UR}{\Omega}}\ ,
\end{equation}
where $\beta \equiv (1/R)\,{\rm d}f/{\rm d}\theta =  
(2\Omega/R)\,\cos\theta$ 
({\it distinct from the ``plasma beta parameter''} $\,\beta_p$).
At the mid-latitudes, the characteristic value of $\beta$ is 
$\beta_0 = \Omega\sqrt{2}/R$.
For our fiducial MSPs, $L_\beta \sim 6\times 10^4 \cm$; here
notably, $L_\beta \sim L_D$.
For slowly rotating RPs, $L_\beta \gtrsim R$. 
This ``conversion of small-scale turbulence into large-scale waves'' 
also gives rise to a more sharply-peaked energy spectrum $\hat{E}(k)$ 
on the low--$k$ side of the spectrum.

\begin{figure}[t]
\centering
\includegraphics[clip, trim=0.0cm 0.0cm 0.0cm -1.0cm, width=7.0cm]{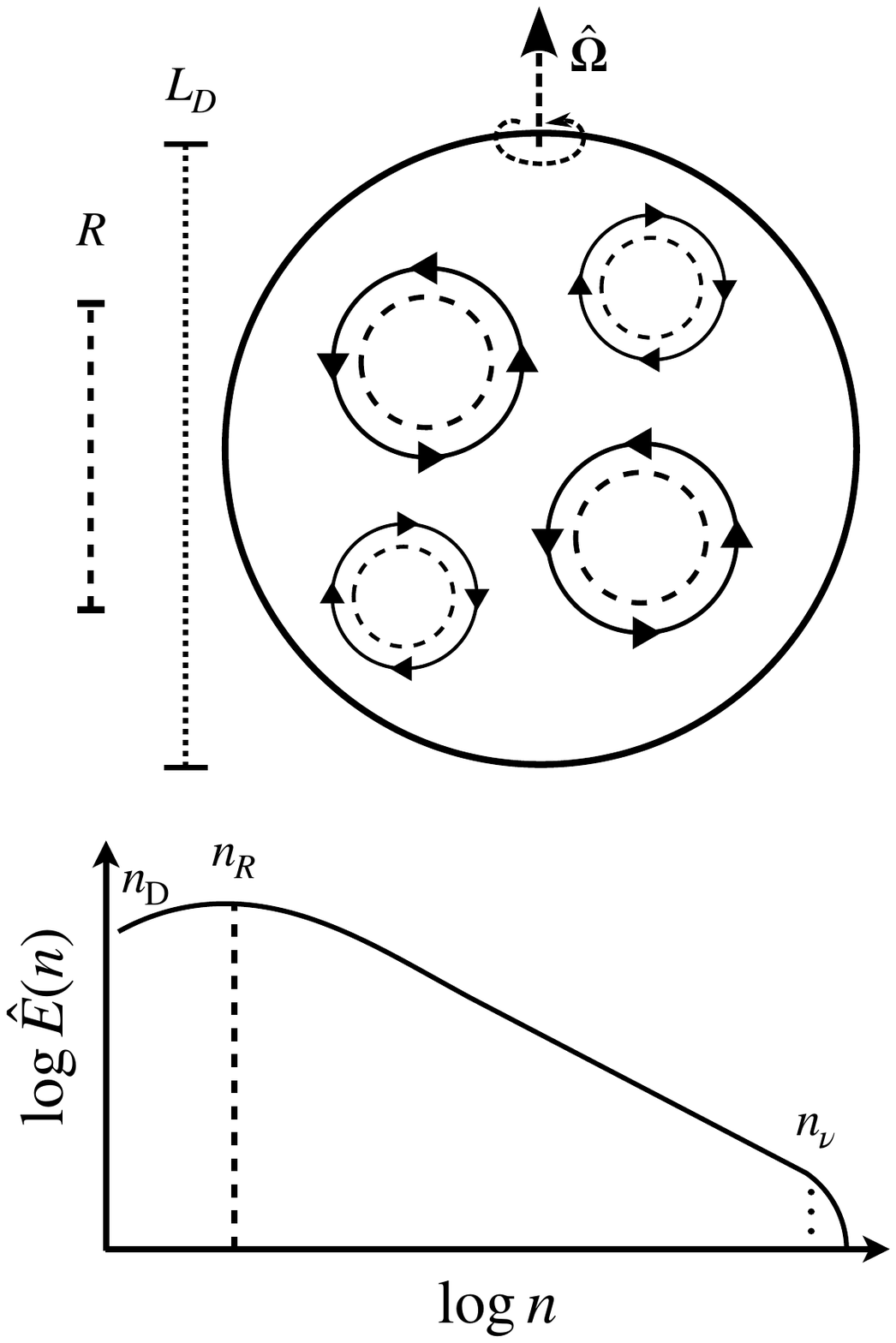}
\caption{\label{fig:slow}
    Illustration of the flow pattern and kinetic energy spectrum for a slowly-rotating, weakly-magnetized neutron star ($R_o,\,\beta_\mathrm{p} \gg 1$).
    Large-scale vortices are expected from the inverse cascade, associated with continuous mergers and growth in the size of the vortices (top panel).
    Accordingly, the energy spectrum $\hat{E}(n)$ is broad (bottom panel); 
    here $n$ is the total wavenumber of a spherical harmonic expansion.
    Three scales are highlighted in the latter: 
    the stellar scale $R$ ($= \pi/n_{R}$),
    Rossby deformation scale, $L_D$ ($= \pi/n_{D}$),
    and dissipation scale $\ell_\nu$ ($= \pi/ n_\nu$). 
}
\end{figure}

Fig.~\ref{fig:slow} presents an example of the HD flow on a neutron 
star when $R_o,\, \beta_\mathrm{p} \gg 1$ and $L_D \rightarrow \infty$.
The resulting turbulent flow is composed of numerous vortices spanning 
a broad range of scales \citep[e.g.,][]{cho1996}.
Faster rotation modifies this picture.
Instead of vortices, the Coriolis force acts as a restoring force in 
the meridional (north--south) direction, suppressing the growth of 
vortices beyond $L_\beta$ scale in that direction \citep{rhines1975}. 
In contrast, no restoring force is offered in the zonal (east--west) 
direction, promoting elongated flows in the zonal 
direction \citep[e.g.,][]{cho1996}.
As the jets form, they sharpen due to nonlinear interactions with 
vortices on both sides of the jet boundary; this induces a 
staircase-like structure of the potential vorticity $q$ and banded appearance of the layer in the meridional 
direction \citep[e.g.,][]{cho1996a,dritschel2011}.

\begin{figure}[t]
\centering
\includegraphics[clip, trim=0.0cm 0.0cm 0.0cm -1.0cm, width=7.0cm]{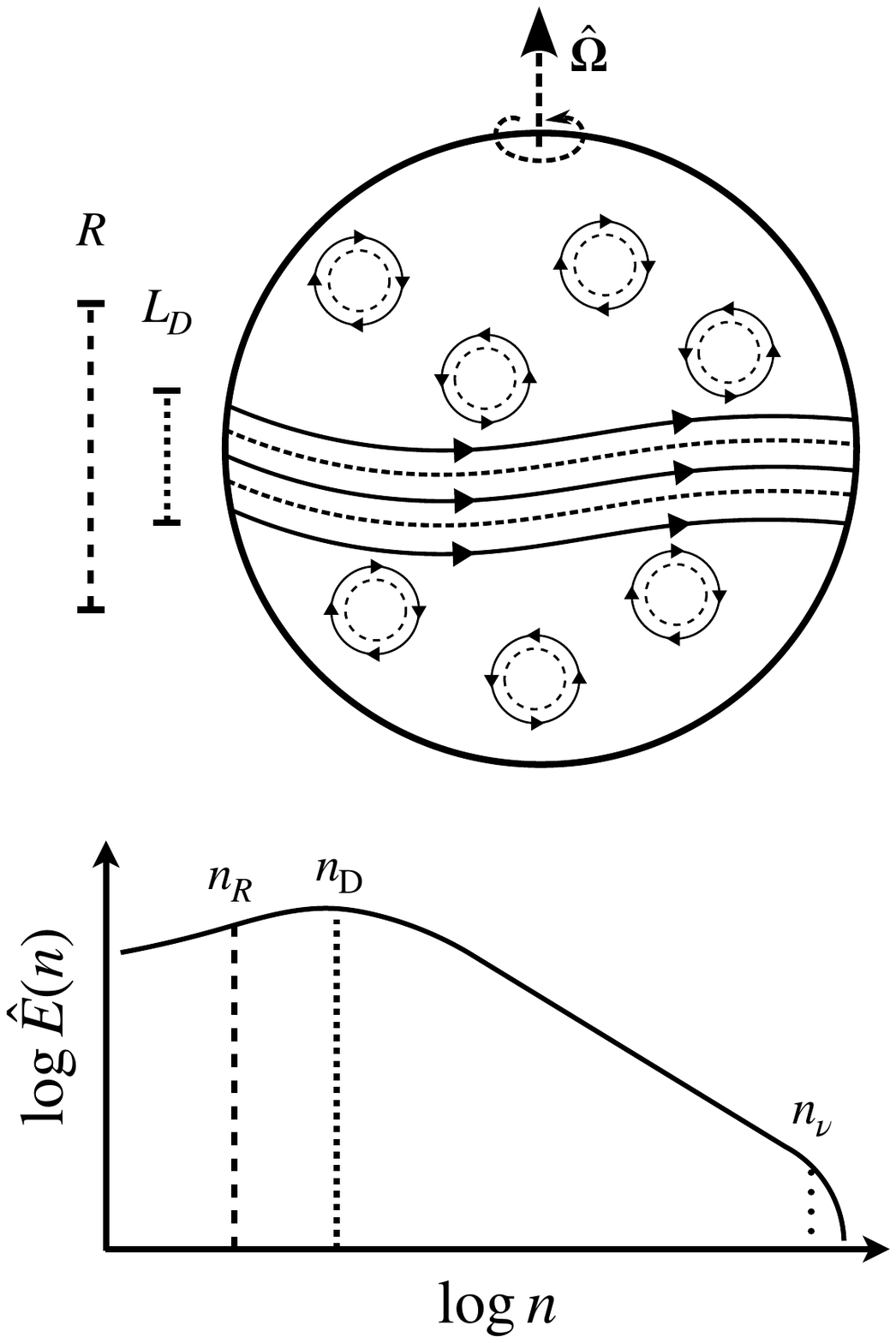}
\caption{\label{fig:fast}
    Illustration of the flow pattern and kinetic energy spectrum for a 
    rapidly-rotating, weakly-magnetized neutron star
    ($R_o \ll 1$ and $\beta_\mathrm{p} \gg 1$).
    Vortices are expected at high latitudes, and zonal jet streams are 
    expected at the equator (top panel).
    Accordingly, the energy spectrum $\hat{E}(n)$ is narrow (bottom panel), compared to the slowly-rotating case.
    The three characteristic scales in Fig.~\ref{fig:slow} are 
    highlighted here as well; notably, $L_D \ll R$. 
    }
\end{figure}

On the sphere, vortex mergers and jet formation can be suppressed at the poles because of the latitudinal variation of the Coriolis acceleration: 
$f$ is weakest and $\beta$ is strongest at the equator---and vice versa at the poles.
The latitudinal dependence of $L_D$, through $f$, also plays a significant role in shaping the global flow pattern via the mediation of mergers \citep{cho1996}.
If $L_D \gtrsim L_\beta$, the number of zonal jets from pole to pole $N_\mathrm{jets}$ is $\sim\!\pi R/ L_\beta$.
For the MSPs, $N_\mathrm{jets} \sim 100$; the actual number of bands is somewhat smaller because of the spherical geometry.  
In contrast, if $L_D \ll L_\beta$, there is a critical latitude beyond which multiple jets and bands cease to form---thus mitigating the band formation; 
see Fig.~\ref{fig:fast}.
Balance of $L_D$ and $L_\beta$ gives a a critical latitude $\theta_{\rm crit}$ of
\begin{equation}
\frac{\sin^2\theta_{\rm crit}}{\cos\theta_{\rm crit}} \approx \theta_{\mathrm{crit}}^2  \sim\ 
\frac{gH}{4UR\Omega}\, .
\end{equation}
Therefore, jets should only appear in a narrow latitudinal width 
$\Delta\theta$ of $\sim\! 2\theta_{\rm crit}$, centered on the equator.
On MSPs, $\Delta\theta \lesssim 60^\circ$.

\vspace{1cm}
\subsection{Flows in Magnetohydrodynamic Regime}\label{sect:mhd}

Turbulence is markedly different in the presence of strong magnetic fields, both internal to the fluid $\bfb$ and external to the fluid $\bfB$.
As already discussed, the coupling between these two fields depends on $\beta_p$.  
Unlike in 2D HD flows, the energy forward cascades in 2D MHD flows---ultimately dissipating at the small scales near $k_\eta$.
In contrast, in 2D MHD flows, the magnetic potential~$a$ (defined by $\bfb = \bfk \times \bfnabla a)$ is the quantity that inverse cascades \citep{iroshnikov1964, kraichnan1965, biskamp1989}. 
For turbulent MHD flows with large magnetic Reynolds numbers, magnetic reconnection can also alter the cascade \citep[e.g.,][]{dong2018}.
Recall that the total potential ${\cal A}$ ($\equiv \langle\, a^2\, \rangle$) is conserved in the MHD case (see Appendix~\ref{app:mhdswe}).
The electric current, $j = |\, c/4\pi\,\bfnabla\times\bfb\,|$\, is the MHD analog of $\zeta$ in HD. 
However, the magnetic field (or current) does not couple to the rotation, since the magnetic induction equation does not depend on Coriolis force.
Therefore, when $\beta_p \lesssim 1$, the resulting vortices do not deform into zonal structures near the equator, as they do when $\beta_p \gg 1$.
Instead, in the former, flows are expected to ``crystallize'' into a lattice of current spikes (magnetic islands or plasmoids), irrespective of the rotation rate \citep{cho2008a}. 
In this case, the flow pattern actually looks similar to that in Fig.~\ref{fig:slow}, but the pattern is induced by the $\bfb$-field rather than the $\bfv$-field.

Strongly magnetized neutron stars, such as magnetars, can also have 
some regions of the atmosphere--ocean that cool down more efficiently 
due to enhanced heat conduction along the $\bfB$-field lines.
Hence, temperature variations in the layers are expected near the 
magnetic poles of the stars, where strong radial field lines are 
present.
This can cause the ocean to freeze \citep[e.g.,][]{heyl1998}.
Such a configuration would require more complex boundary conditions 
as well as the layers to be (horizontally) inhomogeneous.
In this case, the frozen regions can be modeled as solidified ``polar caps''.

\section{Conclusions}\label{sec:conclusions}

Persistent turbulence naturally arises in rotating shallow fluid layers, such as atmospheres and oceans \citep[e.g.,][]{cho1996,cho2008a}.
Here we have used simple scaling estimates and well-understood mechanisms from geophysical and planetary fluid dynamics, to deduce large-scale flow patterns expected for  the atmosphere--ocean layers on neutron stars.
Broadly, neutron star atmosphere--ocean layers can be separated into a thick hydro-like ocean overlayed by a thin MHD atmosphere which itself is enveloped by a very thin, small-$\beta_p$ ``chromosphere''.  
A striking feature of the dynamic layers is long-lived spots (vortices or plasmoids), which appear for any realistic rotation rates of neutron stars (Fig.~\ref{fig:slow}).
Long-lived jets (bands) also emerge on rapidly-rotating neutron stars (Fig.~\ref{fig:fast}).
However, if $\beta_p \lesssim 1$, jet formation is suppressed on the rapid rotators.

Understanding the lateral inhomogeneity as well as the temporal variability of the fluid layers on neutron stars is important because all the observed radiation originates from these layers.
The dynamics in the atmosphere--ocean layers can induce non-uniform, 
large-scale temperature structures on the surface; 
and, this can, for example, bias astrophysical measurements of the star's radius \citep[see, e.g.,][]{nattila2017}.

Our analysis indicates that rapid rotators are the most susceptible to such measurement biases because they have zonally-symmetric but meridionally-varying jet streams that do not induce rotation-phase-dependent flux variations. 
Similarly, persistent background flows can markedly modify the location and size of spots, which are used in pulse profile analysis \citep[e.g.,][]{nattila2018}.  
In contrast, slow-rotators are most likely to exhibit spatial inhomogeneities due to the presence of few, stellar-scale spots, which move chaotically, and lack stationary jets.
Lastly, the flows can lead to enhanced mixing as well as to generation of sharp gradients of chemical species and their sequestration into insulated ``pockets''.
These will affect, for example, the ignition of thermonuclear X-ray 
bursts and propagation of the burning fronts; 
indeed, there is evidence for temporally-changing chemical compositions during X-ray bursts \citep[e.g.,][]{kajava2017}.

\section*{Acknowledgments}
The authors thank Valentin Skoutnev for helpful discussions.
JN is supported by the European Union (ERC, ILLUMINATOR, 101114623). 
Views and opinions expressed are however those of the author(s) only and do not necessarily reflect those of the European Union or the European Research Council. Neither the European Union nor the granting authority can be held responsible for them.
JWS is supported by NASA grant 80NSSC23K0345 and a Simons Foundation Pivot Fellowship.
Support for this work was provided by NASA through the NASA Hubble 
Fellowship grant HST-HF2-51518.001-A awarded by the Space Telescope Science Institute, which is operated by the Association of Universities 
for Research in Astronomy, Incorporated, under NASA contract NAS5-26555.
Part of this work was performed at the Aspen Center for Physics, which 
is supported by National Science Foundation grant PHY-2210452. 

\bibliography{refs}

\begin{thebibliography}{}
\expandafter\ifx\csname natexlab\endcsname\relax\def\natexlab#1{#1}\fi
\providecommand{\url}[1]{\href{#1}{#1}}
\providecommand{\dodoi}[1]{doi:~\href{http://doi.org/#1}{\nolinkurl{#1}}}
\providecommand{\doeprint}[1]{\href{http://ascl.net/#1}{\nolinkurl{http://ascl.net/#1}}}
\providecommand{\doarXiv}[1]{\href{https://arxiv.org/abs/#1}{\nolinkurl{https://arxiv.org/abs/#1}}}

\bibitem[{{Abolmasov} \& {Lipunova}(2022)}]{abolmasov2022}
{Abolmasov}, P., \& {Lipunova}, G. 2022, arXiv e-prints, arXiv:2207.12312, \dodoi{10.48550/arXiv.2207.12312}

\bibitem[{Abolmasov {et~al.}(2020)Abolmasov, N{\"a}ttil{\"a}, \& Poutanen}]{abolmasov2020}
Abolmasov, P., N{\"a}ttil{\"a}, J., \& Poutanen, J. 2020, \aap, 638, A142, \dodoi{10.1051/0004-6361/201936958}

\bibitem[{AlGendy \& Morsink(2014)}]{algendy2014}
AlGendy, M., \& Morsink, S.~M. 2014, \apj, 791, 78, \dodoi{10.1088/0004-637X/791/2/78}

\bibitem[{{Anderson} \& {Itoh}(1975)}]{anderson1975}
{Anderson}, P.~W., \& {Itoh}, N. 1975, \nat, 256, 25, \dodoi{10.1038/256025a0}

\bibitem[{{Batchelor}(1969)}]{batchelor1969}
{Batchelor}, G.~K. 1969, Physics of Fluids, 12, II, \dodoi{10.1063/1.1692443}

\bibitem[{Bildsten(1995)}]{bildsten1995}
Bildsten, L. 1995, \apj, 438, 852, \dodoi{10.1086/175128}

\bibitem[{Biskamp \& Welter(1989)}]{biskamp1989}
Biskamp, D., \& Welter, H. 1989, \phflb, 1, 1964, \dodoi{10.1063/1.859060}

\bibitem[{{Blaes} {et~al.}(1989){Blaes}, {Blandford}, {Goldreich}, \& {Madau}}]{blaes1989}
{Blaes}, O., {Blandford}, R., {Goldreich}, P., \& {Madau}, P. 1989, \apj, 343, 839, \dodoi{10.1086/167754}

\bibitem[{{Braginskii}(1965)}]{braginskii1965}
{Braginskii}, S.~I. 1965, Reviews of Plasma Physics, 1, 205

\bibitem[{{Burrows} \& {Lattimer}(1986)}]{burrows1986}
{Burrows}, A., \& {Lattimer}, J.~M. 1986, \apj, 307, 178, \dodoi{10.1086/164405}

\bibitem[{Cavecchi {et~al.}(2016)Cavecchi, Levin, Watts, \& Braithwaite}]{cavecchi2016}
Cavecchi, Y., Levin, Y., Watts, A.~L., \& Braithwaite, J. 2016, \mnras, 459, 1259, \dodoi{10.1093/mnras/stw728}

\bibitem[{Cavecchi \& Spitkovsky(2019)}]{cavecchi2019}
Cavecchi, Y., \& Spitkovsky, A. 2019, \apj, 882, 142, \dodoi{10.3847/1538-4357/ab3650}

\bibitem[{Cavecchi {et~al.}(2013)Cavecchi, Watts, Braithwaite, \& Levin}]{cavecchi2013}
Cavecchi, Y., Watts, A.~L., Braithwaite, J., \& Levin, Y. 2013, \mnras, 434, 3526, \dodoi{10.1093/mnras/stt1273}

\bibitem[{Cavecchi {et~al.}(2015)Cavecchi, Watts, Levin, \& Braithwaite}]{cavecchi2015}
Cavecchi, Y., Watts, A.~L., Levin, Y., \& Braithwaite, J. 2015, \mnras, 448, 445, \dodoi{10.1093/mnras/stu2764}

\bibitem[{Chambers \& Watts(2020)}]{chambers2020}
Chambers, F. R.~N., \& Watts, A.~L. 2020, \mnras, 491, 6032, \dodoi{10.1093/mnras/stz3449}

\bibitem[{Chamel \& Haensel(2008)}]{chamel2008}
Chamel, N., \& Haensel, P. 2008, \lrir, 11, 10, \dodoi{10.12942/lrr-2008-10}

\bibitem[{{Cho}(2008)}]{cho2008a}
{Cho}, J.~{\relax Y-K}. 2008, Philosophical Transactions of the Royal Society A, 366, 4477, \dodoi{10.1098/rsta.2008.0177}

\bibitem[{{Cho} {et~al.}(2008){Cho}, {Menou}, {Hansen}, \& {Seager}}]{cho2008}
{Cho}, J.~{\relax Y-K}., {Menou}, K., {Hansen}, B. M.~S., \& {Seager}, S. 2008, \apj, 675, 817, \dodoi{10.1086/524718}

\bibitem[{Cho \& Polvani(1996{\natexlab{a}})}]{cho1996}
Cho, J.~{\relax Y-K}., \& Polvani, L.~M. 1996{\natexlab{a}}, \phfl, 8, 1531, \dodoi{10.1063/1.868929}

\bibitem[{Cho \& Polvani(1996{\natexlab{b}})}]{cho1996a}
---. 1996{\natexlab{b}}, \science, 273, 335, \dodoi{10.1126/science.273.5273.335}

\bibitem[{Cho {et~al.}(2019)Cho, Thrastarson, Koskinen, Read, Tobias, Moon, \& Skinner}]{cho2019}
Cho, J.~{\relax Y-K}., Thrastarson, H.~{\relax Th}., Koskinen, T.~T., {et~al.} 2019, in Zonal {Jets}: {Phenomenology}, {Genesis}, and {Physics}, ed. B.~Galperin \& P.~L. Read (Cambridge: Cambridge University Press), 104--116, \dodoi{10.1017/9781107358225.005}

\bibitem[{{Chugunov} \& {Yakovlev}(2005)}]{chugunov2005}
{Chugunov}, A.~I., \& {Yakovlev}, D.~G. 2005, Astronomy Reports, 49, 724, \dodoi{10.1134/1.2045323}

\bibitem[{Cumming \& Bildsten(2000)}]{cumming2000}
Cumming, A., \& Bildsten, L. 2000, \apj, 544, 453, \dodoi{10.1086/317191}

\bibitem[{{Dong} {et~al.}(2018){Dong}, {Wang}, {Huang}, {Comisso}, \& {Bhattacharjee}}]{dong2018}
{Dong}, C., {Wang}, L., {Huang}, Y.-M., {Comisso}, L., \& {Bhattacharjee}, A. 2018, \prl, 121, 165101, \dodoi{10.1103/PhysRevLett.121.165101}

\bibitem[{Dritschel(2004)}]{Dritschel2004ANS}
Dritschel, D. 2004, in Seminar on Recent developments in numerical methods for atmospheric and ocean modelling (Shinfield Park, Reading: ECMWF), 251--263

\bibitem[{Dritschel(1988)}]{dritschel1988}
Dritschel, D.~G. 1988, \jfm, 194, 511, \dodoi{10.1017/S0022112088003088}

\bibitem[{{Dritschel}(1993)}]{dritschel1993b}
{Dritschel}, D.~G. 1993, \phfla, 5, 173, \dodoi{10.1063/1.858802}

\bibitem[{Dritschel \& Scott(2011)}]{dritschel2011}
Dritschel, D.~G., \& Scott, R.~K. 2011, Philosophical Transactions of the Royal Society A: Mathematical, Physical and Engineering Sciences, 369, 754, \dodoi{10.1098/rsta.2010.0306}

\bibitem[{{Dritschel} \& {Tobias}(2023)}]{dritschel2023}
{Dritschel}, D.~G., \& {Tobias}, S.~M. 2023, Journal of Fluid Mechanics, 973, A17, \dodoi{10.1017/jfm.2023.746}

\bibitem[{Eiden {et~al.}(2020)Eiden, Zingale, Harpole, Willcox, Cavecchi, \& Katz}]{eiden2020}
Eiden, K., Zingale, M., Harpole, A., {et~al.} 2020, \apj, 894, 6, \dodoi{10.3847/1538-4357/ab80bc}

\bibitem[{Garcia {et~al.}(2018)Garcia, Chambers, \& Watts}]{garcia2018}
Garcia, F., Chambers, F. R.~N., \& Watts, A.~L. 2018, \phrvf, 3, 123501, \dodoi{10.1103/PhysRevFluids.3.123501}

\bibitem[{Gill(1982)}]{gill1982}
Gill, A.~E. 1982, Atmosphere-ocean dynamics (New York, NY: Academic Press)

\bibitem[{Gilman(2000)}]{gilman2000}
Gilman, P.~A. 2000, \apj, 544, L79, \dodoi{10.1086/317291}

\bibitem[{Gudmundsson {et~al.}(1982)Gudmundsson, Pethick, \& Epstein}]{gudmundsson1982}
Gudmundsson, E.~H., Pethick, C.~J., \& Epstein, R.~I. 1982, \apj, 259, L19, \dodoi{10.1086/183840}

\bibitem[{Haensel {et~al.}(2007)Haensel, Potekhin, \& Yakovlev}]{haensel2007}
Haensel, P., Potekhin, A.~Y., \& Yakovlev, D.~G. 2007, Neutron {{Stars}} 1 : {{Equation}} of {{State}} and {{Structure}} (New York, NY: Springer), \dodoi{10.1007/978-0-387-47301-7}

\bibitem[{Harpole {et~al.}(2021)Harpole, Ford, Eiden, Zingale, Willcox, Cavecchi, \& Katz}]{harpole2021}
Harpole, A., Ford, N.~M., Eiden, K., {et~al.} 2021, \apj, 912, 36, \dodoi{10.3847/1538-4357/abee87}

\bibitem[{Heng \& Spitkovsky(2009)}]{heng2009}
Heng, K., \& Spitkovsky, A. 2009, \apj, 703, 1819, \dodoi{10.1088/0004-637X/703/2/1819}

\bibitem[{Heyl(2004)}]{heyl2004}
Heyl, J.~S. 2004, \apj, 600, 939, \dodoi{10.1086/379966}

\bibitem[{Heyl \& Hernquist(1998)}]{heyl1998}
Heyl, J.~S., \& Hernquist, L. 1998, \mnras, 300, 599, \dodoi{10.1046/j.1365-8711.1998.01885.x}

\bibitem[{Inogamov \& Sunyaev(1999)}]{inogamov1999}
Inogamov, N.~A., \& Sunyaev, R.~A. 1999, \astl, 25, 269

\bibitem[{Inogamov \& Sunyaev(2010)}]{inogamov2010}
---. 2010, \astl, 36, 848, \dodoi{10.1134/S1063773710120029}

\bibitem[{Iroshnikov(1964)}]{iroshnikov1964}
Iroshnikov, P.~S. 1964, \sva, 7, 566

\bibitem[{Kajava {et~al.}(2017)Kajava, N{\"a}ttil{\"a}, Poutanen, Cumming, Suleimanov, \& Kuulkers}]{kajava2017}
Kajava, J. J.~E., N{\"a}ttil{\"a}, J., Poutanen, J., {et~al.} 2017, \mnras, 464, L6, \dodoi{10.1093/mnrasl/slw167}

\bibitem[{Klimachkov \& Petrosyan(2016)}]{klimachkov2016}
Klimachkov, D.~A., \& Petrosyan, A.~S. 2016, \jetp, 123, 520, \dodoi{10.1134/S1063776116070098}

\bibitem[{Kraichnan(1965)}]{kraichnan1965}
Kraichnan, R.~H. 1965, \phfl, 8, 1385, \dodoi{10.1063/1.1761412}

\bibitem[{Kraichnan(1967)}]{kraichnan1967}
---. 1967, \phfl, 10, 1417, \dodoi{10.1063/1.1762301}

\bibitem[{Lee(2004)}]{lee2004}
Lee, U. 2004, \apj, 600, 914, \dodoi{10.1086/380122}

\bibitem[{{Leith}(1968)}]{leith1968}
{Leith}, C.~E. 1968, Physics of Fluids, 11, 671, \dodoi{10.1063/1.1691968}

\bibitem[{Lewin {et~al.}(1993)Lewin, {van Paradijs}, \& Taam}]{lewin1993}
Lewin, W. H.~G., {van Paradijs}, J., \& Taam, R.~E. 1993, \ssrv, 62, 223, \dodoi{10.1007/BF00196124}

\bibitem[{McWilliams(1984)}]{mcwilliams1984}
McWilliams, J.~C. 1984, \jfm, 146, 21, \dodoi{10.1017/S0022112084001750}

\bibitem[{{N{\"a}ttil{\"a}} \& {Kajava}(2022)}]{nattila2022c}
{N{\"a}ttil{\"a}}, J., \& {Kajava}, J. J.~E. 2022, in Handbook of X-ray and Gamma-ray Astrophysics, ed. C. Bambi \& A. Santangelo (Springer Living Reference Work), 30, \dodoi{10.1007/978-981-16-4544-0_105-1}

\bibitem[{N{\"a}ttil{\"a} {et~al.}(2017)N{\"a}ttil{\"a}, Miller, Steiner, Kajava, Suleimanov, \& Poutanen}]{nattila2017}
N{\"a}ttil{\"a}, J., Miller, M.~C., Steiner, A.~W., {et~al.} 2017, \aap, 608, A31, \dodoi{10.1051/0004-6361/201731082}

\bibitem[{N{\"a}ttil{\"a} \& Pihajoki(2018)}]{nattila2018}
N{\"a}ttil{\"a}, J., \& Pihajoki, P. 2018, \aap, 615, A50, \dodoi{10.1051/0004-6361/201630261}

\bibitem[{\"Ozel \& Freire(2016)}]{Ozel:2016oaf}
\"Ozel, F., \& Freire, P. 2016, Ann. Rev. Astron. Astrophys., 54, 401, \dodoi{10.1146/annurev-astro-081915-023322}

\bibitem[{Pedlosky(1979)}]{pedlosky1979}
Pedlosky, J. 1979, Geophysical fluid dynamics (New York, NY: Springer), \dodoi{10.1007/978-1-4684-0071-7}

\bibitem[{Polvani {et~al.}(1989)Polvani, Flierl, \& Zabusky}]{polvani1989}
Polvani, L.~M., Flierl, G.~R., \& Zabusky, N.~J. 1989, \phfla, 1, 181, \dodoi{10.1063/1.857485}

\bibitem[{{Potekhin}(1999)}]{potekhin1999}
{Potekhin}, A.~Y. 1999, \aap, 351, 787, \dodoi{10.48550/arXiv.astro-ph/9909100}

\bibitem[{{Potekhin} \& {Chabrier}(2000)}]{potekhin2000}
{Potekhin}, A.~Y., \& {Chabrier}, G. 2000, \pre, 62, 8554, \dodoi{10.1103/PhysRevE.62.8554}

\bibitem[{Potekhin {et~al.}(1997)Potekhin, Chabrier, \& Yakovlev}]{potekhin1997}
Potekhin, A.~Y., Chabrier, G., \& Yakovlev, D.~G. 1997, \aap, 323, 415

\bibitem[{Potekhin {et~al.}(2015)Potekhin, Pons, \& Page}]{potekhin2015}
Potekhin, A.~Y., Pons, J.~A., \& Page, D. 2015, \ssrv, 191, 239, \dodoi{10.1007/s11214-015-0180-9}

\bibitem[{{Potekhin} {et~al.}(2003){Potekhin}, {Yakovlev}, {Chabrier}, \& {Gnedin}}]{potekhin2003}
{Potekhin}, A.~Y., {Yakovlev}, D.~G., {Chabrier}, G., \& {Gnedin}, O.~Y. 2003, \apj, 594, 404, \dodoi{10.1086/376900}

\bibitem[{Reinaud \& Dritschel(2018)}]{reinaud2018}
Reinaud, J.~N., \& Dritschel, D.~G. 2018, \jfm, 848, 388, \dodoi{10.1017/jfm.2018.367}

\bibitem[{Rhines(1975)}]{rhines1975}
Rhines, P.~B. 1975, \jfm, 69, 417, \dodoi{10.1017/S0022112075001504}

\bibitem[{Ripa(1993)}]{ripa1993}
Ripa, P. 1993, \gapfd, 70, 85, \dodoi{10.1080/03091929308203588}

\bibitem[{Scott(2010)}]{scott2010}
Scott, R.~K. 2010, in {{IUTAM Symposium}} on {{Turbulence}} in the {{Atmosphere}} and {{Oceans}}, ed. D.~Dritschel, Vol.~28 ({Springer, Dordrecht}), 243--252, \dodoi{10.1007/978-94-007-0360-5_20}

\bibitem[{Spitkovsky {et~al.}(2002)Spitkovsky, Levin, \& Ushomirsky}]{spitkovsky2002}
Spitkovsky, A., Levin, Y., \& Ushomirsky, G. 2002, \apj, 566, 1018, \dodoi{10.1086/338040}

\bibitem[{{Spitzer}(1962)}]{spitzer1962}
{Spitzer}, L. 1962, {Physics of Fully Ionized Gases} (Hoboken, NJ: John Wiley and Sons)

\bibitem[{{Strohmayer} \& {Bildsten}(2006)}]{strohmayer2006}
{Strohmayer}, T., \& {Bildsten}, L. 2006, in Compact Sterllar X-ray Sources, Vol.~39 (Cambridge, UK: Cambridge University Press), 113--156, \dodoi{10.1017/CBO9780511536281.004}

\bibitem[{Strohmayer(2001)}]{strohmayer2001}
Strohmayer, T.~E. 2001, Advances in Space Research, 28, 511, \dodoi{10.1016/S0273-1177(01)00422-7}

\bibitem[{{Sturrock}(1971)}]{sturrock1971}
{Sturrock}, P.~A. 1971, \apj, 164, 529, \dodoi{10.1086/150865}

\bibitem[{{Sullivan} {et~al.}(2023){Sullivan}, {Alves}, {Spence}, {Leite}, {Veske}, {Bartos}, {M{\'a}rka}, \& {M{\'a}rka}}]{sullivan2022}
{Sullivan}, A.~G., {Alves}, L. M.~B., {Spence}, G.~O., {et~al.} 2023, \mnras, 520, 6173, \dodoi{10.1093/mnras/stad389}

\bibitem[{{Ventura} \& {Potekhin}(2001)}]{ventura2001}
{Ventura}, J., \& {Potekhin}, A. 2001, in The Neutron Star - Black Hole Connection, ed. C.~{Kouveliotou}, J.~{Ventura}, \& E.~{van den Heuvel}, Vol. 567, 393, \dodoi{10.48550/arXiv.astro-ph/0104003}

\bibitem[{Watts(2012)}]{watts2012}
Watts, A.~L. 2012, \araa, 50, 609, \dodoi{10.1146/annurev-astro-040312-132617}

\end{thebibliography}
\bibliographystyle{aasjournal}


\begin{appendix}

\section{Magnetohydrodynamic Shallow-Water Equations}\label{app:mhdswe}

The magnetohydrodynamic shallow-water equations (MHDSWE) describe the dynamics of a thin, isopycnal layer of ``magnetized'' fluid \citep{gilman2000}.
The motions of the layer are two-dimensional (2D), $\bfx = (x,y)\, \in\, \mathbb{R}^2$\ (Fig.~\ref{fig:layer}).
The MHDSWE describe the time evolution of velocity $\bfv(\bfx, t)$, free surface geopotential $\phi(\bfx, t)$, and magnetic field $\bfb(\bfx, t)$.
The dimensional form of the MDHSWE are:
\begin{align}
    \frac{\partial \bfv}{\partial t} + \bfv\! \cdot\! \bfnabla\, \bfv\ &=\ 
    -\bfnabla\, \phi  - \bff \times \bfv  + \bfb\! \cdot\! \bfnabla\, \bfb  + 
    \nu\, \bfnabla^2 \bfv, \label{eq:mhd_dim1}\\
    \frac{\partial \phi}{\partial t} +  \bfv\! \cdot\! \bfnabla \phi\ &=\ 
    -\phi\,\bfnabla\! \cdot\! \bfv, \label{eq:mhd_dim2} \\
    \frac{\partial \bfb}{\partial t} + \bfv\! \cdot\! \bfnabla \bfb \ &=\ 
    \bfb\!\cdot\! \bfnabla\, \bfv + \eta\, \nabla^2 \bfb, \label{eq:mhd_dim3}
\end{align}
where 
$\bff = 2\Omega \cos\theta\, \bfk$ is the Coriolis parameter,
$\Omega$ is the angular velocity of the rotating frame,
$\theta$ is the latitude,
$\bfk$ is the unit vector in the local vertical direction,
$\nu$ is the viscosity coefficient, and
$\eta$ is the magnetic diffusivity coefficient;
$\phi = g h(\bfx,t)$, where $g$ is the gravitational acceleration (assumed to be constant across the thin layer) and $h$ is the surface height of the layer.  
By assumption, the full three-dimensional (3D) velocity field as well as the 3D magnetic field are solenoidal:
\begin{align}\label{eq:div}
\bfnabla \cdot \bfv + \frac{\partial w}{\partial z}   \ = & \ \ 0 \\
\bfnabla \cdot \bfb + \frac{\partial b_z}{\partial z} \ = & \ \ 0, 
\end{align}
where $w$ and $b_z$ are the vertical velocity and magnetic field components, respectively.

We scale Eqs.~\eqref{eq:mhd_dim1}--\eqref{eq:mhd_dim3} via a characteristic horizontal advection speed $U$,  horizontal length $L$, and Alfv\'en speed $v_A = |\bfb|/\sqrt{4\pi \rho}$:
\begin{align}
\begin{split}\label{eq:mhdwe_long}
\bfv\,  \rightarrow\, & \ U \bfv\, , \qquad\qquad\qquad
\frac{\partial}{\partial t}\,  \rightarrow \, 
\frac{U}{L}\, \frac{\partial}{\partial t}\, , \qquad\qquad\qquad
\bfnabla\,  \rightarrow\,  \frac{1}{L}\, \bfnabla\, , \qquad \\
h\,  \rightarrow\, & \ H h\, , \qquad\qquad\qquad\ 
\bfb\,   \rightarrow\,  \frac{1}{\sqrt{4\pi \rho}} \bfb\, , 
\qquad\qquad\quad\ \,
\bfj\,  \rightarrow\,  \frac{c}{4\pi}\, \bfj\ , \qquad \\
\end{split}
\end{align}
where the last scaling is anticipating the appearance of the electric current density field $\bfj$.
This results in the non-dimensional form of the MHDSWE:
\begin{align}
    \frac{\partial \bfv}{\partial t} + \bfv\! \cdot\! \bfnabla\, \bfv \ &=\ 
    -\frac{g H}{U^2}\, \bfnabla\, h  
    - \frac{2\Omega L}{U}\, \bfft \times \bfv  
    + \frac{b^2}{4\pi\rho U^2}\,\bfb\! \cdot\! \bfnabla\, \bfb  
    + \frac{\nu}{U L}\, \bfnabla^2\, \bfv \nonumber \\
    &= \ \ 
    -\frac{1}{F_r^{\ 2}}\,\nabla\, h 
    - \frac{ 1}{R_o}\,\bfft \times \bfv 
    + \frac{1}{\beta_p^{\ 2}}\bfb\! \cdot\! \bfnabla\, \bfb 
    + \frac{1}{R_e}\,\bfnabla^2\, \bfv \label{eq:mhdswe1}\\
    \frac{\partial h}{\partial t} +  \bfv \cdot \bfnabla h \ &=\ 
    -h\,\bfnabla \cdot \bfv \qquad 
    \label{eq:mhdswe2}\\
    \frac{\partial \bfb}{\partial t} + \bfv\! \cdot\! \bfnabla\, \bfb \ &=\ 
    \bfb\! \cdot\! \bfnabla\, \bfv + \frac{\eta}{L U}\, \nabla^2\, \bfb
    \nonumber \\ 
    &=\ \ \bfb\! \cdot\! \bfnabla\, \bfv 
    + \frac{1}{R_m}\, \nabla^2\, \bfb\ , \label{eq:mhdswe3}
\end{align}
where $\mathbf{f} \rightarrow 2\Omega\,\bfft$ scaling has been used so that $\bfft = \sin\theta\,\bfk$. 
The resulting dimensionless parameters, 
\{$F_r$, $R_o$, $\beta_p$, $R_e$, $R_m$\}, are discussed in the main text.
The above equations correspond to Eqs.~\eqref{eq:mhdswe_momentum}--\eqref{eq:mhdswe_induction}.

When $\bfb = 0$, Eqs.~(\ref{eq:mhdswe1})--(\ref{eq:mhdswe3}) 
possess several useful global invariants in the absence of dissipation ($R_e \rightarrow \infty$): 
\begin{itemize}[topsep=8pt,itemsep=2pt,leftmargin=3\parindent]
    \item total height,\ $H_\Sigma\ =\ \Big\langle\, h\, \Big\rangle$
    \item total energy,\ $\mathcal{E}\ =\ \Big\langle\, \frac{1}{2} \left(\bfv^2 + gh \right)\,h\,\Big\rangle$
    \item total potential enstrophy,\ $\mathcal{Z}\ =\ \Big\langle\,\frac{1}{2}\, q^2\,\Big\rangle$\, ,
\end{itemize}
where $\langle\,\cdot\,\rangle$ denotes the domain average, and  
\begin{equation}\label{eq:pv}
    q\ \equiv\ \frac{\zeta + f}{h}
\end{equation}
is the potential vorticity.
In Eq.~(\ref{eq:pv}), $\zeta \equiv \bfk\cdot\nabla\times\bfv$ is the (relative) vorticity, and $f \equiv 2\Omega \sin\theta$ is the Coriolis parameter; 
in the latter, $\Omega = 2\pi/P$ is the stellar rotation rate, and $\theta$ is the latitude.
Importantly, $q$ is a material invariant---i.e., $\rmD q/{\rm D}t = 0$.  
However, when $\bfb \ne 0$, this invariance is broken.
In this case, a new invariant arises---the magnetic potential $a(\bfx,t)$, defined by $\bfb = \bfk \times \bfnabla a$; 
thus, $a$ is the magnetic analog of the velocity streamfunction.
In the absence of resistivity ($R_m \rightarrow \infty$), some useful invariants of the more general MHDSWE are: 
\begin{itemize}[topsep=8pt,itemsep=2pt,leftmargin=3\parindent]
    \item total height,\ $H_\Sigma\ =\ \Big\langle\, h\, \Big\rangle$
    \item total energy,\ $\mathcal{E}'\ =\ \Big\langle\, \frac{1}{2}\,\left( \bfv^2 + gh + \bfb^2\right)\,h\, \Big\rangle$\, .
\end{itemize}
Additional invariants are known for pure-2D MHD systems, which include:
\begin{itemize}[topsep=8pt,itemsep=2pt,leftmargin=3\parindent]    
    \item total potential,\ $\mathscr{A}\ =\ \Big\langle\, a^2\, \Big\rangle$
    \item total cross-helicity,\ $\mathscr{H}_C\ =\ \Big\langle\, \bfv\! \cdot\! \bfb \,\Big\rangle$\, .
\end{itemize}

\newpage
\section{Vortex Dynamics}\label{app:vortex}

\begin{figure*}[ht]
\centering
\includegraphics[clip, trim=0.0cm 25.0cm 0.0cm 0.0cm, width=0.9\textwidth]{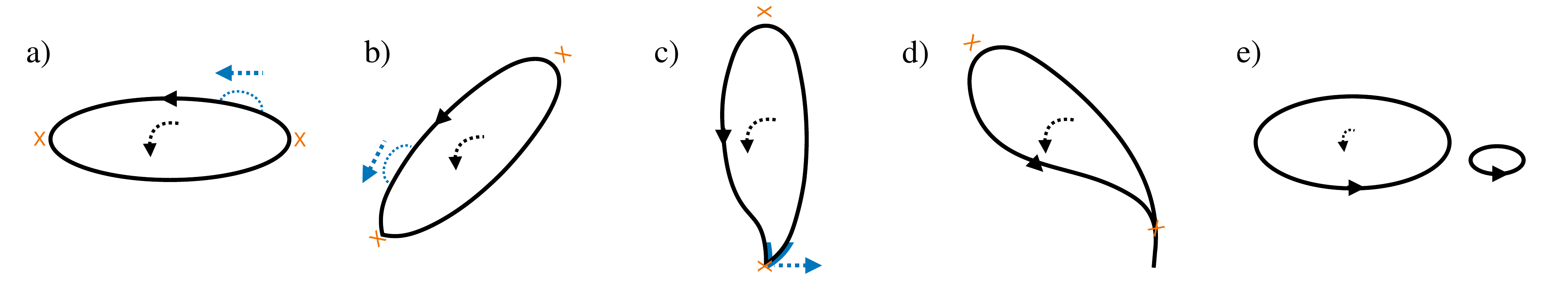}
\caption{\label{fig:vortex}
    Vortex filamentation and splitting.
    a) Material contour of a (counter-clockwise) rotating ellipsoidal patch of vorticity with a Kelvin wave disturbance propagating along its edge (blue dotted line) is shown; the flow around the vortex has two hyperbolic critical points (orange crosses).
    b,c) As the vortex rotates, its tip (superposition of the ellipsoidal vortex and the crest of the Kelvin wave) crosses the critical point. 
    d) Material contour is distorted into a filament by strong velocity shear.
    e) Contour is disconnected (pinched off) at a narrow point along the filament and reconnects into two separate vortices or one vortex if the second one does not survive further straining and subsequent dissipation.
}
\end{figure*}

\begin{figure*}[th]
\centering
\includegraphics[clip, trim=0.0cm 27.0cm 0.0cm 0.0cm, width=0.9\textwidth]{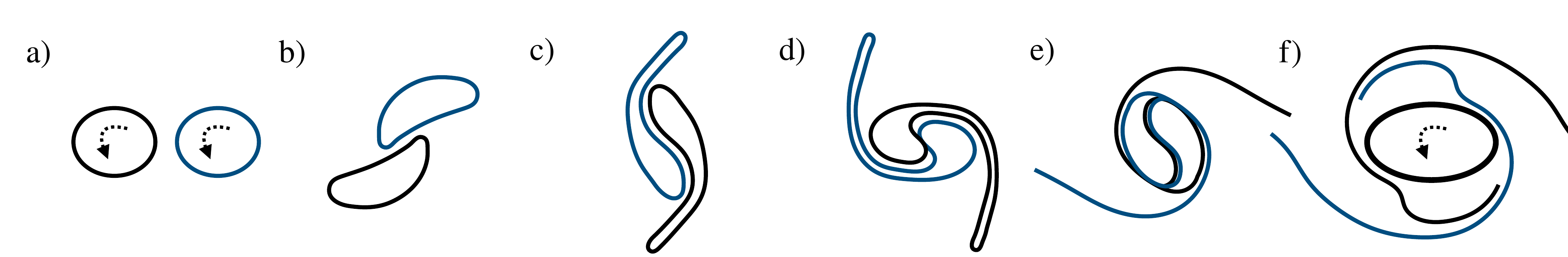}
\caption{\label{fig:vortex_merger2}
    Vortex merging.
    a) Material contours of two (counter-clockwise) rotating vortices are shown.
    b) As the vortices coalesce, their shape begins to deform due to the strain caused by the flow around the companion.
    c) The straining stretches the far ends of the vortices into  filaments while the close ends curl up to form a separate central region.
    d) The vorticity contours in the central region mix and form a (larger) merger remnant.
    e) The tails stretch into thin filaments that disconnect from the central region.
    f) The merging process generates a large vortex and (mostly filamentary) vorticity debris.
}     
\end{figure*}

\begin{figure*}[th]
\centering
\includegraphics[clip, trim=0.0cm 27.0cm 0.0cm 0.0cm, width=0.9\textwidth]{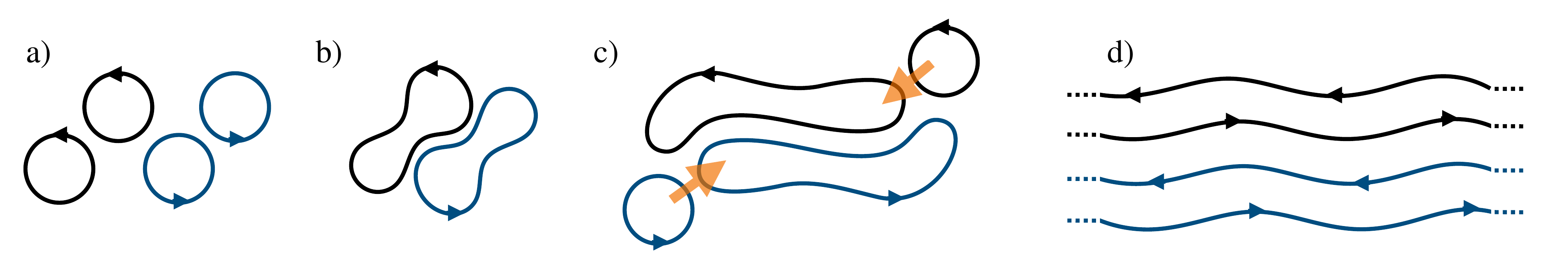}
\caption{\label{fig:vortex_banding}
    Vortex banding.
    a) A group of multiple vortices located close to the equator is shown.
    Some vortices rotate counter-clockwise (black curves) and others clockwise (blue curves).
    b) Vortex mergers occur predominantly along the meridional (East--West; here left--right) direction because the Coriolis force balances the growth in the latitudinal (North--South; here up--down) direction.
    c) The elongated flow structures grow in length by subsequent vortex mergers.
    d) Eventually, the resulting flow forms zonal jet streams that wrap around the whole star.
    The boundaries of the streams sharpen due to additional nonlinear effects.
}
\bigskip
\bigskip          
\end{figure*}

Schematic visualizations summarizing the vortex filamentation (Fig.~\ref{fig:vortex}), merging (Fig.~\ref{fig:vortex_merger2}), and banding (Fig.~\ref{fig:vortex_banding}) mechanisms are presented here.

\bigskip
\bigskip
\bigskip


\end{appendix}
\end{document}